\documentclass[aps,superscriptaddress,tightenlines,eqsecnum,nofootinbib]{revtex4}
\usepackage{graphicx,amssymb,amsbsy,bm,amsmath}

\usepackage{hyperref}

\newcommand{\vx}{\ensuremath{\vec{x}}}
\newcommand{\vk}{\ensuremath{\vec{k}}}

\newcommand{\be}{\begin{equation}}
\newcommand{\ee}{\end{equation}}
\newcommand{\bea}{\begin{eqnarray}}
\newcommand{\eea}{\end{eqnarray}}

\begin{document}
\title{Particle abundance in a thermal plasma: \\
quantum kinetics vs. Boltzmann equation}
\author{D. Boyanovsky}
\email{boyan@pitt.edu--corresponding author} \affiliation{Department
of Physics and Astronomy, University of Pittsburgh, Pittsburgh,
Pennsylvania 15260, USA}
\author{K. Davey}
\email{kdavey@ku.edu} \affiliation{Department of Physics and
Astronomy, University of Pittsburgh, Pittsburgh, Pennsylvania 15260,
USA}
\author{C. M. Ho}
\email{cmho@phyast.pitt.edu} \affiliation{Department of Physics and
Astronomy, University of Pittsburgh, Pittsburgh, Pennsylvania 15260,
USA}
\date{\today}
\begin{abstract}
We study the abundance of a particle species in a thermalized plasma
by introducing a quantum kinetic description based on the
non-equilibrium effective action. A stochastic interpretation of
quantum kinetics in terms of a Langevin equation emerges naturally.
 We consider a particle species that is stable in the vacuum and interacts with \emph{heavier}
particles that constitute a thermal bath in equilibrium. Asymptotic
theory suggests a  definition of a fully renormalized single
particle distribution function. Its real time dynamics is completely
determined by the non-equilibrium effective action which furnishes a
Dyson-like resummation of the perturbative expansion. The
distribution function reaches thermal equilibrium on a time scale
$\sim 1/2\,\Gamma_k(T)$ with $\Gamma_k(T)$ being the
\emph{quasiparticle} relaxation rate. The equilibrium distribution
function depends on the full spectral density as a consequence the
fluctuation-dissipation relation. Such dependence leads to off-shell
contributions to the particle abundance. A specific model of a
bosonic field $\Phi$ in interaction with two \emph{heavier} bosonic
fields $\chi_{1,2}$ is studied. The decay of the heaviest particle
and its recombination lead to a width of the spectral function for
the particle $\Phi$ and to off-shell corrections to the abundance.
We find substantial departures from the Bose-Einstein result both in
the high temperature and the low temperature but high momentum
region. In the latter the abundance is exponentially suppressed but
larger than the Bose-Einstein result. We obtain the Boltzmann
equation in renormalized perturbation theory and highlight the
origin of the differences. Cosmological consequences are discussed:
we argue that the corrections to the abundance of  cold dark matter
candidates are observationally negligible and that recombination
erases any possible spectral distortions of the CMB. However we
expect that the  enhancement  at high temperature may be important
for baryogenesis.
\end{abstract}

\pacs{98.80.Cq;11.10.Wx;05.70.Ln}

\maketitle

\section{Introduction}\label{sec:intro}
Phenomena out of equilibrium played a fundamental role in the early
Universe: during  phase transitions, baryogenesis, nucleosynthesis,
recombination,  particle production, annihilation and freeze out of
relic particles, some of which could be dark matter
candidates\cite{kolbturner,bernstein,dodelson}. Of the many
different non-equilibrium processes, particle production,
annihilation and freeze-out and
baryogenesis\cite{kolbturner,buchmuller} are non-equilibrium kinetic
processes which are mainly studied via the Boltzmann
equation\cite{kolbturner,bernstein,dodelson}.

The Boltzmann kinetic equation is also the main approach to study
equilibration, thermalization and abundance of a species in a
plasma. A thorough formulation of \emph{semiclassical }kinetic
theory in an expanding Friedmann-Robertson-Walker cosmology is given
in ref.\cite{bernstein}.

 However the Boltzmann equation is a classical equation for the
\emph{distribution function} with an inhomogeneity determined by
collision terms which are computed with the S-matrix formulation
of quantum field theory. The collision term in the Boltzmann
equation is obtained from the transition \emph{probability} per
unit time extracted from the asymptotic long time limit of the
transition matrix element. This is tantamount to implementing
Fermi's golden rule. Potential quantum interference and memory
effects are completely ignored in this approach. Furthermore a
single particle distribution function, the main ingredient in the
Boltzmann equation, is usually defined via some coarse graining
procedure. All of these shortcomings of the usual semiclassical
Boltzmann equations when extrapolated to the realm of temperatures
and density in the Early Universe, suggest that in order to
provide a reliable understanding of such delicate processes such
as baryo and leptogenesis a full quantum field theory treatment of
kinetics may be required\cite{buchmuller}.

One of the basic predictions of the Boltzmann equation is that the
local thermodynamic equilibrium solution for the abundance of a
particle species is determined by the Bose-Einstein or Fermi-Dirac
distribution functions, hence exponentially suppressed at low
temperatures (in absence of a chemical potential).

This basic prediction has recently been challenged in a series of
articles\cite{yoshimura} wherein a surprising result is obtained:
the abundance of heavy particles with masses much larger than the
temperature is \emph{not} exponentially suppressed as the
Boltzmann equation predicts but the suppression is a \emph{power
law}. Such result, if correct, can have important consequences for
the relic abundance of cold dark matter candidates.

This result, however,  has been criticized and scrutinized in
detail by several authors\cite{srednicki,jia,pietroni} who
concluded that it is a consequence of the definition of the
particle number introduced in ref.\cite{yoshimura}. The definition
of the \emph{total} number of  particles proposed in
\cite{yoshimura} is based on the non-interacting Hamiltonian for
the heavy particle divided by its mass plus counterterms,  which
purportedly account for renormalization effects. The results of
references\cite{srednicki,jia,pietroni} point out the inherent
ambiguity in separating the contribution to the energy density
from the particle and that of the bath and the interaction. The
ambiguity in the separation of the different contributions to the
energy has been studied thoroughly in these references  in
particular exactly solvable models\cite{srednicki}, effective
field theory\cite{jia} or a consistent treatment of
renormalization effects\cite{pietroni}.

Understanding the limitations of and corrections to the Boltzmann
kinetic description and potential departures from the predicted
abundances is important for a deeper assessment of possible
mechanisms of baryogenesis as well as for the relic abundance of
cold dark matter candidates. In the case of baryogenesis, the
applicability and reliability  of Boltzmann kinetics in the
conditions of temperature and density that prevailed in the early
Universe warrants a critical reassessment\cite{buchmuller}.
Refinements of the usual Boltzmann equation have been proposed in
the literature\cite{jakovac}.

While the work in refs.\cite{srednicki,jia,pietroni} has clarified
the shortcomings of the definition of the \emph{total} particle
number proposed in\cite{yoshimura} explaining the origin of the
power law suppression as a consequence of the ambiguity in this
definition, what is missing from this discussion is a suitable
definition of a \emph{distribution function} and its real time
evolution. The Boltzmann equation is a local differential equation
that determines the dynamics of the \emph{single particle
distribution function}. Therefore in order to clearly assess
potential corrections to the equilibrium solutions of the familiar
Boltzmann equation a suitable distribution function and its
dynamical evolution must be understood.

The definition of the distribution function both in non-relativistic
many body theory\cite{baym} as well as in relativistic quantum field
theory\cite{malfliet,calzetta} is typically based on a Wigner
transform of a two point correlation function, which is not
 manifestly positive semidefinite. Usual derivations of the  Boltzmann kinetic
equation invoke gradient expansions or quasiparticle (on-shell)
approximations which lead to Markovian dynamics. Alternative
derivations of the kinetic equations\cite{boyaboltz} which
explicitly implement real time perturbation theory often invoke a
long time limit and Fermi's Golden rule which enforces energy
conservation in the kinetic equation. This is also the case in the
dynamical renormalization group approach to quantum kinetics
advocated in ref.\cite{drg} although this latter method allows one
to systematically include off-shell corrections. Whichever method of
derivation of the kinetic equation is used, the first step is to
\emph{define} a single particle distribution function.

Any definition of the distribution function of particles that
\emph{decay} in the vacuum (resonances) is fraught with ambiguities
because the spectral representation of such particles is not a sharp
delta function but typically a Breit-Wigner distribution. Since
these particles decay even in \emph{vacuum} and do not exist as
asymptotic states any definition of an operator that ``counts''
these particles will unavoidably be ambiguous.

In this work, we circumvent this ambiguity by  focusing on the
study of the quantum kinetics and  equilibration dynamics of the
distribution functions of particles that are \emph{stable} at zero
temperature associated with a field $\Phi$. Stable physical
particles are asymptotic states which can be measured and a
distribution function for the single particle physical states can
be introduced according to the basic assumptions  of asymptotic
theory. While our ultimate goal is to find a quantum kinetic
description for phenomena in the early Universe, in this article
we focus on a study in Minkowski space-time as a first step
towards that goal.

\vspace{1mm}

\textbf{Goals and methods:}

\vspace{1mm}

In this article we provide a framework for non-equilibrium quantum
kinetics beyond the usual Boltzmann equation. This non-equilibrium
formulation  includes off-shell and non-Markovian (memory)
processes which are not accounted for in the semiclassical
Boltzmann equation and result in modifications of the equilibrium
abundances. We focus on the case of a scalar field $\Phi$ coupled
to other heavier fields for a wide variety of relevant interacting
quantum field theories. Here we consider that the \emph{heavier}
fields  constitute a thermal bath in equilibrium. In order to
study the thermalization of the $\Phi$ particle as well as the
time evolution of its distribution function we consider the case
in which the field $\Phi$ is coupled to the thermal bath at some
initial time $t_i$. We then obtain the \emph{non-equilibrium}
effective action for the field $\Phi$ by integrating out the
degrees of freedom of the thermal bath  to lowest order in the
coupling of the field $\Phi$ to the heavy sector but in principle
to \emph{all orders} in the couplings of the heavy fields amongst
themselves.

At zero temperature  the $\Phi$-particles are stable because they
are the lightest, therefore they are manifest as asymptotic
states. Hence according to asymptotic theory we introduce a
definition of an interpolating number operator that counts these
particles, for example as those measured by a detector in a
collision experiment in the vacuum. At finite temperature the
distribution function is the expectation value of this
interpolating operator in the statistical ensemble. The real time
evolution of this distribution function is completely determined
by the  non-equilibrium effective action and its asymptotic long
time limit determines the abundance of the physical particles
$\Phi$ in the thermal plasma. The non-equilibrium approach
introduced here, borrows from the seminal work on quantum Brownian
motion\cite{schwinger,feyver,leggett,grabert} which is adapted to
quantum field theory.

After the discussion of the general case, we introduce a specific
model in which the scalar field $\Phi$ associated with the stable
particle couples to two heavier bosonic fields which constitute the
thermal bath. At lowest order in the coupling we find that the
$\Phi$ particle despite being the lightest, acquires a width in the
medium as a consequence of the two body decay of the heavier
particle and its recombination in the plasma. These processes result
in a broadening of its spectral function and corrections to its
equilibrium abundance.

\vspace{1mm}

 \textbf{Brief summary of results:}

 \begin{itemize}
 \item{We obtain the non-equilibrium effective action for a field $\Phi$ coupled to other \emph{heavier}
 fields by integrating out the latter to lowest order in their coupling to the field $\Phi$ but
  in principle to \emph{all orders}  in the couplings amongst themselves. The heavy fields are taken to be in thermal equilibrium
 and therefore provide a thermal ``bath'' for the $\Phi$ field.
  The resulting non-equilibrium effective action can be interpreted as a generating functional
 of a \emph{stochastic} field theory in which the (integrated out) heavy fields introduce a Gaussian but colored noise and a non-Markovian
 self-energy (dissipative) kernel.   }

 \item{We introduce a definition of the single particle distribution function in the general case of a particle that is \emph{stable} in the
 vacuum. Stable physical particles are asymptotic states which can be measured by a detector. In accordance with the results of asymptotic theory,
 we introduce a fully renormalized interpolating number operator whose expectation value in the non-equilibrium state (density matrix) is
 identified with  the single particle distribution
 function. The time evolution of this distribution function is determined by the non-equilibrium effective action and is completely
 specified by the solution of a stochastic Langevin equation with a memory kernel and a Gaussian stochastic noise. The properties of the
 memory kernel are related to the spectrum of the noise by a generalized fluctuation dissipation relation. We argue that the time evolution of the
 distribution function is a result of  a  Dyson  resummation of the perturbative expansion provided by the non-equilibrium effective action. The
 single particle distribution function becomes insensitive to the initial conditions at time scales longer than the ``quasiparticle'' relaxation
 time and its asymptotic long time limit  describes a thermalized state.   }

 \item{ A specific example is studied in detail. This is  a model  of Bosonic scalar fields with a coupling $g\,\Phi \,\chi_1\,\chi_2$ with the masses of the ``bath'' fields
 $\chi_{1,2}$ obeying the hierarchy  $M_1 > M_2 >>M_{\Phi}$. In this case the particles associated with the field $\Phi$ are stable in the vacuum.
 However, at  finite temperature the particle $\Phi$ \emph{acquires a  width } from the two-body decay and recombination process
 $\chi_1 \leftrightarrow \Phi+\chi_2$. We study the
 approach to thermal equilibrium of the single $\Phi$ particle distribution function whose asymptotic long time limit yields their equilibrium abundance
  in the bath. We find that the equilibrium abundance is always
  \emph{larger} than that predicted by the Bose-Einstein distribution. The enhancement is more significant at high temperatures, as
well as at low temperatures but large momenta. The departure from
the Bose-Einstein result is a distinct consequence  of off-shell
support of the spectral function of the $\Phi$ field in the
plasma. }

 \item{We derive the usual quantum kinetic Boltzmann equation in renormalized perturbation theory  up to the same order in the coupling to the
 bath as the  non-equilibrium effective action. This derivation highlights the neglect of memory and correlations in the usual Boltzmann
 equation. We contrast  its prediction for the equilibrium abundance, the usual Bose-Einstein distribution, to that from the full quantum kinetic equation
 with memory and off-shell   contributions. This  direct comparison leads to the conclusion that  memory and off-shell phenomena result in substantial
  corrections to the   equilibrium  abundances that are not captured by the Boltzmann equation.   }

  \item{ We conclude that potential corrections to the abundance of
  cold dark matter candidates as well as distortions of the cosmic
  microwave background post recombination are negligible observationally, but
  substantial corrections in a high temperature plasma may be
  important for baryogenesis.}

\end{itemize}

\vspace{1mm}

The article is organized as follows: in section (\ref{sec:noneLeff})
we introduce the general form of the interacting quantum field
theories considered and develop the formulation in terms of the
non-equilibrium effective action. The effective action is obtained
to lowest order in the coupling of the field $\Phi$ to the heavier
fields (the bath) and in principle to \emph{all orders} in the
coupling of the bath fields amongst themselves. We show that a
stochastic formulation in terms of a Langevin equation emerges
naturally. In section (\ref{sec:count}) we introduce the definition
of the fully renormalized interpolating number operator and the
single particle distribution function based on asymptotic theory.
The time evolution of this distribution function is completely
determined by the solution of the stochastic Langevin equation.

In section (\ref{sec:model}) we study a specific model in which the
$\Phi$ field is coupled to two heavy scalar fields with a coupling
$g\,\Phi\,\chi_1\,\chi_2$. This interacting quantum field theory
provides an excellent testing ground and highlights the main
conceptual results. We study the dynamics of the distribution
function for the $\Phi$ particle up to one loop order. The
asymptotic distribution function is studied for a wide range of
parameters allowing to extract fairly general conclusions whose
validity goes beyond this specific model. In particular we analyze
in detail how off-shell effects  result in \emph{large} corrections
to the usual Bose-Einstein equilibrium abundance. In section
(\ref{sec:boltzkin}) we obtain the usual Boltzmann quantum kinetic
equation and highlight the main assumptions implicit in its
derivation. We contrast the predictions for the asymptotic abundance
between the non-equilibrium kinetic formulation and that of the
usual quantum kinetic Boltzmann equation, highlighting that memory
and off-shell effects are responsible for the differences in the
predictions. Our conclusions and a  discussion on the cosmological
consequences are presented in section (\ref{sec:conc}). An appendix
is devoted to the explicit calculation of the self-energy in the
specific example studied.

\section{General formulation: the non-equilibrium effective
action}\label{sec:noneLeff}

 We focus on the description of the
dynamics of the relaxation of the occupation number of a scalar
field $\Phi$ which is in interaction with other fields either
fermionic or bosonic, collectively written as $\chi_i$, with a
Lagrangian density of the form
\begin{equation}\label{lagra}
{\cal L}[\Phi(x),\chi(x)]= {\cal L}_{0,\Phi}[\Phi(x)]+{\cal
L}_{\chi_i}[\chi_i(x)]+g\Phi{\cal O}[\chi_i(x)]
\end{equation}
where ${\cal O}[\chi_i]$ stands for an operator non-linear  in the
fields $\chi_i$ and ${\mathcal L}_{0,\Phi}$ is the free field
Lagrangian density for the field $\Phi$ but ${\cal
L}_{\chi_i}[\chi_i(x)]$ is the full Lagrangian for the fields $\chi$
including interactions amongst themselves. This general form
describes several relevant cases:
\begin{itemize}

\item{Interacting scalars, for example the linear sigma model in
the broken symmetry phase. The interaction between the massive
scalar and the Goldstone bosons is of the form $\sigma \pi^2$. In
this article we focus on the case of a  trilinear interaction of
the form $\Phi \sum_{ij}g_{ij}\chi_i \chi_j$ where the fields
$\chi_{1,2}$ have masses larger than that of the $\Phi$ field. }

\item{ A Yukawa theory with $\chi$ being fermionic fields and
$\Phi$ a scalar field, with interaction $\Phi \bar{\Psi}\Psi$.
This could be generalized to a chiral model.}

\item{ A gauge theory in which $\Phi$ is the gauge field and
$\chi$ is either a complex scalar or fermion fields, the interaction
being of the form $A^{\mu}J_{\mu}$ with $J_{\mu}$ being a bilinear
of the fields. In particular this approach has been recently
implemented to study photon production from a quark gluon plasma in
local thermal equilibrium\cite{boyphoton}. This case is particularly
relevant for assessing potential distortions in the spectrum of the
cosmic microwave background.}

\item{ Another possible realization of this situation could be the case in which $\Phi$ is a neutrino
field in interaction with leptons and (or) quarks which constitute a
thermal or dense plasma.}

\item{The case of a self-interacting scalar field in which one
mode say with wave vector $k$ is singled out as the ``system'' and
the other modes are treated as a ``bath''.}

\end{itemize}

 In all of these cases the fields $\chi_i$ are treated as
a bath in equilibrium assuming that the bath fields are sufficiently
strongly coupled so as to guarantee their thermal equilibration.
These fields will be ``integrated out'' yielding a reduced density
matrix for the field $\Phi$ in terms of an effective real-time
functional, known as  the influence functional\cite{feyver} in the
theory of quantum brownian motion. The reduced density matrix can be
represented by a path integral in terms of the non-equilibrium
effective action that includes the influence functional.  This
method has been used extensively to study quantum brownian
motion\cite{feyver,leggett,grabert} and for preliminary studies of
quantum kinetics in the simpler case of a particle coupled linearly
to a bath of harmonic oscillators\cite{boyalamo,yoshimura}.

The models can be generalized further by considering that the
interaction between $\Phi$ and $\chi$ is also polynomial in $\Phi$.
However, in this article we will consider the simpler case described
by (\ref{lagra}) since it describes a broad range of physically
relevant cases,  and as  will be discussed below this case already
reveals a wealth of novel phenomena. As we will discuss in detail
below most of the relevant phenomena can be highlighted within this
wide variety of models and  most  of the results  will be seen to be
fairly general.

The relaxation of the distribution function   is  an initial value
problem, therefore we propose the initial density matrix at a time
$t_i$ to be of the form
\begin{equation}
\hat{\rho}(t_i) = \hat{\rho}_{\Phi,i} \otimes
\hat{\rho}_{\chi_i,i} \label{inidensmtx}
\end{equation}

The initial density matrix of the $\chi_i$ fields will be taken to
describe state in thermal equilibrium at a temperature
$T=1/\beta$, namely

\begin{equation}\label{rhochi}
\hat{\rho}_{\chi} = e^{-\beta\,H_{\chi}}
\end{equation}

\noindent where $H_{\chi_i}(\chi_i)$ is the  Hamiltonian for the
fields $\chi_i$. We will now refer collectively to the set of
fields $\chi_i$ simply as $\chi$ to avoid cluttering of indices.

In the field basis the matrix elements of $\hat{\rho}_{\Phi,i}$
are given by
\begin{equation}
\langle \Phi |\hat{\rho}_{\Phi,i} | \Phi'\rangle =
\rho_{\Phi,i}(\Phi ;\Phi')
\end{equation}
The density matrix for $\Phi$ will represent an initial out of
equilibrium state.

  The physical situation described by this initial state is that
  of a  field (or fields)  in thermal equilibrium at a temperature
  $T=1/\beta$, namely a heat bath,  which is put in contact with another system, here represented by the field $\Phi$.
   Once the system  and bath are put in contact their mutual interaction will
  eventually lead to a state of thermal equilibrium.  The goal is to study the
  relaxation of the field $\Phi$ towards equilibrium with the
  ``bath''. The initial density matrix of the field $\Phi$ will
  describe a state with few quanta (or the vacuum) initially.

  The real time evolution of this initial uncorrelated state will
  introduce transient evolution, however the long time behavior
  will be insensitive to this initial transient. Furthermore,
  we point out that it is important to study the initial transient
  stage for the following reason. As a particle $\Phi$ propagates
  in the medium it will be screened or dressed by the excitations
  in the medium and it will propagate as a ``quasiparticle''. Its
  distribution function will be shown to become insensitive to the
  initial conditions on time scales larger than the
  ``quasiparticle'' relaxation time.

The strategy is to integrate out the $\chi$ fields therefore
obtaining the reduced time dependent density matrix for the field
$\Phi$, and the non-equilibrium influence functional for this
field. Once we obtain the reduced density matrix for the field
$\Phi$ we can compute expectation values or correlation functions
of this field. We will focus on studying the time evolution of the
distribution function, or particle number to be defined below.

The time evolution of the initial density matrix is given by

\begin{equation}\label{rhooft}
\hat{\rho}(t_f)= e^{-iH(t_f-t_i)}\hat{\rho}(t_i)e^{iH(t_f-t_i)}
\end{equation}

Where the total Hamiltonian $H$ is given by
\begin{equation}\label{hami}
H=H_{\Phi}(\Phi) + H_{\chi}(\chi)+H_I(\Phi,\chi)
\end{equation}

The calculation of correlation functions is facilitated by
introducing currents coupled to the different fields. Furthermore
since each time evolution operator in eqn. (\ref{rhooft}) will be
represented as a path integral, we introduce different sources for
 forward and backward time evolution operators, referred to as
 $J^{+},J^{-}$ respectively.  The
forward and backward time evolution operators in presence of
sources are $U(t_f,t_i;J^+)$, $U^{-1}(t_f,t_i,J^{-})$
respectively.

We will only study  correlation functions of the $\Phi$ field,
therefore we carry out the trace over the $\chi$ degrees of freedom.
Since the currents $J^{\pm}$ allow us to obtain the correlation
functions for any arbitrary time by simple variational derivatives
with respect to these sources, we can take $t_f\rightarrow \infty$
without loss of generality.

The non-equilibrium generating functional is  given by

\be \label{noneqgen}{\cal Z}[j^+,j^-] =
\mathrm{Tr}U(\infty,t_i;J^+)\hat{\rho}(t_i)U^{-1}(\infty,t_i,J^{-})
\ee

Where  $J^{\pm}$ stand collectively for all the sources coupled to
different fields. Functional derivatives with respect to the
sources $J^+$ generate the time ordered correlation functions,
those with respect to $J^-$ generate the anti-time ordered
correlation functions and mixed functional derivatives with
respect to $J^+,J^-$ generate mixed correlation functions. Each
one of the time evolution operators in the generating functional
(\ref{noneqgen}) can be written in terms of a path integral: the
time evolution operator $U(\infty,t_i;J^+)$ involves a path
integral \emph{forward} in time from $t_i$ to $t=\infty$ in
presence of sources $J^+$, while the inverse time evolution
operator $U^{-1}(\infty,t_i,J^{-})$ involves a path integral
\emph{backwards} in time from $t=\infty$ back to $t_i$ in presence
of sources $J^-$. Finally the equilibrium density matrix for the
bath $e^{-\beta\,H_{\chi}} $ can be written as a path integral
along imaginary time with sources $J^{\beta}$. Therefore the path
integral form of the generating functional (\ref{noneqgen}) is
given by

\be {\cal Z}[j^+,j^-] =
 \int D\Phi_i \int  D\Phi'_i\,
\rho_{\Phi,i}(\Phi_i;\Phi'_i) \int
 {\cal D}\Phi^{\pm}
\int {\cal D} \chi^{\pm} {\cal D}
\chi^{\beta}e^{iS[\Phi^{\pm},\chi^{\pm};J^{\pm}_{\Phi};J^{\pm}_{\chi}]}
 \label{pathint}\ee

 \noindent with the boundary conditions $\Phi^+(\vec
 {x},t_i)=\Phi_i(\vec{x})\,;\,\Phi^-(\vec
 {x},t_i)=\Phi^{'}_i(\vec{x})$.

\begin{figure}[ht!]
\begin{center}
\includegraphics[height=2in,width=4in,keepaspectratio=true]{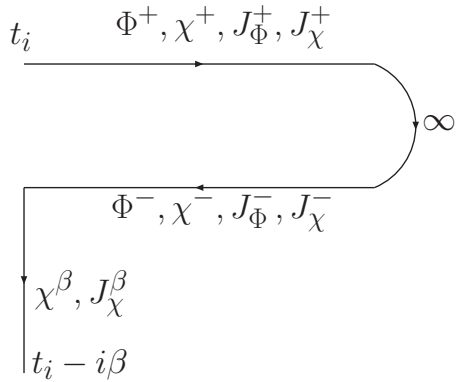}
\caption{Contour in time for the non-equilibrium path integral
representation. } \label{fig:ctp}
\end{center}
\end{figure}

The non-equilibrium  action is given by
 \bea
S[\Phi^{\pm},\chi^{\pm};J^{\pm}_{\Phi};J^{\pm}_{\chi}] & = &
\int_{t_i}^{\infty}dt
d^3x\,\left[\mathcal{L}_{0,\Phi}(\Phi^+)+J^+_{\Phi}\Phi^++h\Phi^+
-\mathcal{L}_{0,\Phi}(\Phi^-)-J^-_{\Phi}\Phi^- - h\Phi^- \right]
+\nonumber
\\ &&
\int_{\mathcal{C}}d^4x\Bigg\{\mathcal{L}_{\chi}(\chi)+ J_{\chi}\chi
+g\,\Phi\,\mathcal{O}[\chi] \Bigg\} \label{noneqlagradens} \eea

\noindent where $\mathcal{C}$ describes a contour in the complex
time plane as follows: along the forward branch $(t_i,+\infty)$
 the fields and sources are $\Phi^{+},\chi^+,J^+_{\chi}$,
along the backward branch  $(\infty, t_i)$  the fields and sources
are $\Phi^{-},\chi^-,J^-_{\chi}$ and along the Euclidean branch
$(t_i, t_i-i\beta)$  the fields and sources are $\Phi=0;
\chi^{\beta},J^{\beta}_{\chi}$. Along the Euclidean branch the
interaction term vanishes since the initial density matrix for the
field $\chi$ is assumed to be that of thermal equilibrium. The
contour is depicted in fig. (\ref{fig:ctp})

The linear term $h\Phi^{\pm}$ is a counterterm that will be required
to cancel the linear terms (tadpole) in $\Phi^{\pm}$ in the
non-equilibrium effective action. This issue will be discussed below
when we obtain the non-equilibrium effective action for the field
$\Phi$ after integrating out the field(s) $\chi$.

The trace over the degrees of freedom of the $\chi$ field with the
initial equilibrium  density matrix,  entail periodic (for bosons)
or antiperiodic (for fermions) boundary conditions for $\chi$
along the contour $\mathcal{C}$. However,  the boundary conditions
on the path integrals for the field $\Phi$ are given by

\be \Phi^+(\vec x,t=\infty) =\Phi^-(\vec x,t=\infty)\label{endbc}\ee

\noindent and
\begin{equation}
\Phi^+(\vec x,t=t_i)= \Phi_i(\vec x)  \; \; \; ; \; \; \;
\Phi^-(\vec x,t=t_i)=\Phi'_i(\vec x) \label{condsfi}
\end{equation}

The reason for the different path integrations is that whereas the
$\chi$ field is traced over with an initial thermal density matrix
(since it is taken as the ``bath''), the initial density matrix for
the $\Phi$ field will be specified later as part of the initial
value problem. The path integral over $\chi$ leads to the influence
functional for $\Phi^{\pm}$\cite{feyver}.

\subsection{Tracing over the ``bath'' degrees of freedom}

As far as the path integrals over the bath degrees of freedom $\chi$
is concerned the fields $\Phi^\pm$ are simply c-number sources. The
contour path integral

\be Z[\Phi^\pm]= \int {\cal D} \chi^{\pm} {\cal D}
\chi^{\beta}e^{i\int_{\mathcal{C}}d^4x\Bigg\{\mathcal{L}_{\chi}(\chi)+
J_{\chi}\chi +g\,\Phi\,\mathcal{O}[\chi] \Bigg\} }
 \label{pathintchi}\ee

\noindent is the generating functional of correlation functions of
the field $\chi$ in presence of external c-number sources
$\Phi^{\pm}$ (the sources $J^{\pm}_{\chi}$ generate the correlation
functions via functional derivatives and are set to zero at the end
of the calculation), namely

\be\int {\cal D} \chi^{\pm} {\cal D}
\chi^{\beta}e^{i\int_{\mathcal{C}}d^4x\Big\{\mathcal{L}_{\chi}(\chi)+
g \Phi\,\mathcal{O}[\chi] \Big\} } = \Big\langle e^{ig
\int_{\mathcal{C}}d^4x \Phi\,\mathcal{O}[\chi]}
\Big\rangle_{\chi}\,Z[0]. \label{aver}\ee

Note that the expectation value in the right hand side of eqn.
(\ref{aver}) is in the equilibrium density matrix of the field
$\chi$. The path integral can be carried out in perturbation theory
and the result exponentiated to yield the effective action as
follows

\be \Big\langle e^{ig \int_{\mathcal{C}}d^4x
\Phi\,\mathcal{O}[\chi]} \Big\rangle_{\chi} =  1+ ig
 \int_{\mathcal{C}}d^4x
 \Phi(x)\,\Big\langle\mathcal{O}[\chi](x)\Big\rangle_{\chi}+
 \frac{(ig)^2}{2}\int_{\mathcal{C}}d^4x
 \int_{\mathcal{C}}d^4x'\Phi(x)\Phi(x')\Big\langle
 \mathcal{O}[\chi](x)\mathcal{O}[\chi](x')\Big\rangle_{\chi}+\mathcal{O}(g^3)
 \ee

This the usual expansion of the exponential of the connected
correlation functions, where this series is identified with

\be \Big\langle e^{ig \int_{\mathcal{C}}d^4x
\Phi\,\mathcal{O}[\chi]} \Big\rangle_{\chi} =
e^{i\,L_{if}[\Phi^+,\Phi^-] }\;,
 \ee
\noindent and  where the \emph{influence functional}\cite{feyver}
$L_{if}[\Phi^+,\Phi^-]$ is given by the following expression

\be L_{if}[\Phi^+,\Phi^-] = g\int_{\mathcal{C}}d^4x
 \Phi(x)\,\langle\mathcal{O}[\chi](x)\rangle_{\chi}
 +i\frac{g^2}{2}\int_{\mathcal{C}}d^4x
 \int_{\mathcal{C}}d^4x'\Phi(x)\Phi(x')\langle
 \mathcal{O}[\chi](x)\mathcal{O}[\chi](x')\rangle_{\chi,con}+\mathcal{O}(g^3)\label{Lif}\ee

In detail, the integrals along the contour $\mathcal{C}$ stand for
the following:

\be \int_{\mathcal{C}}d^4x
 \Phi(x)\,\langle\mathcal{O}[\chi](x)\rangle_{\chi} = \int d^3x
 \int_{t_i}^{\infty}dt\left[\Phi^+(\vec{x},t)\langle\mathcal{O}[\chi^+](x)\rangle_{\chi}-
 \Phi^-(\vec{x},t)\langle\mathcal{O}[\chi^-](x)\rangle_{\chi}\right]
 \ee

\bea \int_{\mathcal{C}}d^4x
 \int_{\mathcal{C}}d^4x'\Phi(x)\Phi(x')\langle
 \mathcal{O}[\chi](x)\mathcal{O}[\chi](x')\rangle_{\chi,con} = &&\int
 d^3x \int_{t_i}^{\infty}dt \int
 d^3x' \int_{t_i}^{\infty}dt' \Big[ \Phi^+(x)\Phi^+(x')\langle
 \mathcal{O}[\chi^+](x)\mathcal{O}[\chi^+](x')\rangle_{\chi,con}
 \nonumber\\+&&
 \Phi^-(x)\Phi^-(x')\langle
 \mathcal{O}[\chi^-](x)\mathcal{O}[\chi^-](x')\rangle_{\chi,con}\nonumber\\
 -&& \Phi^+(x)\Phi^-(x')\langle
 \mathcal{O}[\chi^+](x)\mathcal{O}[\chi^-](x')\rangle_{\chi,con}\nonumber \\ -&&
 \Phi^-(x)\Phi^+(x')\langle
 \mathcal{O}[\chi^-](x)\mathcal{O}[\chi^+](x')\rangle_{\chi,con}\Big]\eea

Since the expectation values above are computed in a thermal
equilibrium translational invariant density matrix, it is
convenient to  introduce the spatial Fourier transform of the
composite operator ${\cal O}$ in a spatial volume $V$ as
\begin{equation}
{\cal O}_{\vec k}(t) = \frac{1}{\sqrt{V}} \int d^3x e^{i \vec k
\cdot \vec x}
 {\cal O}[\chi(\vec x,t)]
\label{spatialFT}
\end{equation}
 in terms of which we obtain
following the correlation functions
\begin{eqnarray}
&&\langle {\cal O}_{\vec k}(t)\rangle  =  \langle {\cal O}^+_{\vec
k}(t)\rangle = \langle {\cal O}^-_{\vec k}(t)\rangle = \mathrm{Tr}\,
e^{-\beta\,H_{\chi}}\, {O}_{\vec k}(t)\label{expec}\\&& \langle
{\cal O}_{\vec k}(t) {\cal O}_{-\vec k}(t')\rangle  =  \langle
\mathcal {O}^-_{\vec k}(t) {\cal O}^+_{-\vec k}(t')\rangle =
\mathrm{Tr}\,{\cal O}_{-\vec k}(t')\, e^{-\beta\,H_{\chi}}\,
{O}_{\vec k}(t)= \mathcal{G}^>_k(t-t') =  {\cal
G}^{-+}_k(t,t')\label{ggreat} \\&& \langle {\cal O}_{-\vec k}(t')
{\cal O}_{\vec k}(t)\rangle  = \langle {\cal O}^+_{\vec k}(t) {\cal
O}^{-}_{-\vec k}(t')\rangle= \mathrm{Tr}\,{O}_{\vec k}(t) \,
e^{-\beta\,H_{\chi}}\,  {\cal O}_{-\vec k}(t') ={\cal G}^<_k(t-t') =
{\cal G}^{+-}_k(t,t')= {\cal G}^{-+}_k(t',t)\label{lesser} \\&&
\langle T {\cal O}_{\vec k}(t) {\cal O}_{-\vec k}(t')\rangle  =
{\cal G}^>_k(t-t')\Theta(t-t')+ {\cal G}^<_k(t-t')\Theta(t'-t)=
{\cal G}^{++}_k(t,t') \label{timeordered} \\&& \langle \widetilde{T}
{\cal O}_{\vec k}(t) {\cal O}_{-\vec k}(t')\rangle =  {\cal
G}^>_k(t-t')\Theta(t'-t)+ {\cal G}^<_k(t-t')\Theta(t-t') = {\cal
G}^{--}_k(t,t')\label{antitimeordered}
\end{eqnarray}

The time evolution of the operators is determined by the
Heisenberg picture of $H_{\chi}$, namely ${\cal O}_{\vec k}(t)=
e^{iH_{\chi}(t-t_i)}{\cal O}_{\vec k}(t_i)e^{-iH_{\chi}(t-t_i)}$.
Because the density matrix for the bath is in equilibrium, the
correlation functions above are solely functions of the time
difference. These correlation functions are computed
\emph{exactly} to \emph{all orders} in the  couplings of the bath
fields amongst themselves.

These correlation functions are not independent, but obey

\begin{equation}\label{const}
\mathcal{ G}^{++}_k(t,t') + \mathcal{
G}^{--}_k(t,t')-\mathcal{G}^{-+}_k(t,t')-\mathcal{G}^{+-}_k(t,t')=0
\end{equation}

The non-equilibrium effective action is given by

\be\label{noneqeff} L_{eff}[\Phi^+,\Phi^-] = \int_{t_i}^{\infty}dt
d^3x\,\left[\mathcal{L}_{0,\Phi}(\Phi^+)+h\Phi^+
-\mathcal{L}_{0,\Phi}(\Phi^-) - h\Phi^- \right] +
L_{if}[\Phi^+,\Phi^-]\ee

\noindent where we have set the sources $J^\pm$ for the fields
$\Phi^\pm$ to zero.

The choice of  counterterm

\be h = -\langle {\cal O}(\vx,t) \rangle \label{counter}\ee

\noindent cancels the  terms linear in $\Phi^\pm$ (tadpole) in the
non-equilibrium effective action.

In what follows we take $t_i =0$ without loss of generality since
(i) for $t > t_i$ the total Hamiltonian is time independent and the
correlations will be solely functions of $t-t_i$, and (ii) we will
be ultimately interested in the limit $t \gg t_i$ when all transient
phenomena has relaxed. In terms of the spatial Fourier transform of
the fields $\Phi^\pm$ defined as in eqn. (\ref{spatialFT}) we find

\begin{eqnarray}\label{influfunc}
iL_{eff}[\Phi^+,\Phi^-] & = & \sum_{\vec k}\Bigg\{ \frac{i}{2}
\int_0^{\infty} dt \left[\dot{\Phi}^+_{\vec k}(t)\dot{\Phi}^+_{-\vec
k}(t)- (k^2+m^2)\Phi^+_{\vec k}(t)\Phi^+_{-\vec k}(t)
 -\dot{\Phi}^-_{\vec k}(t)\dot{\Phi}^-_{-\vec k}(t)+(k^2+m^2)\Phi^-_{\vec k}(t)\Phi^-_{-\vec k}(t) \right]   \nonumber \\
&   &  - \frac{g^2}{2} \int_0^{\infty} dt \int_0^{\infty} dt' \left[
\Phi^+_{\vec k}(t) {\cal G}^{++}_k(t,t')\Phi^+_{-\vec k}(t')+
\Phi^-_{\vec k}(t){\cal G}^{--}_k(t,t') \Phi^-_{-\vec k}(t') \right.
\nonumber \\ && \left. -\Phi^+_{\vec k}(t){\cal
G}^{+-}_k(t,t')\Phi^-_{-\vec k}(t')- \Phi^-_{\vec k}(t){\cal
G}^{-+}_k(t,t')\Phi^+_{-\vec k}(t')\right] \Bigg\}
\end{eqnarray}

\noindent where all the time integrations are in the interval $0
\leq t \leq \infty$.

A similar program has been used recently to study the relaxation of
scalar fields\cite{yokoyama} as well as the photon production from a
quark gluon plasma in thermal equilibrium\cite{boyphoton}.

\subsection{ Stochastic description: generalized Langevin equation. }

  As it will become clear below,
it is more convenient to introduce the Wigner center of mass and
relative variables \be \Psi(\vec x,t)   =   \frac{1}{2}
\left(\Phi^+(\vec x,t) + \Phi^-(\vec x,t) \right) \; \; ; \; \;
R(\vec x,t) = \left(\Phi^+(\vec x,t) - \Phi^-(\vec x,t) \right)
\label{wigvars} \ee

\noindent  and the Wigner transform of the initial density matrix
for the $\Phi$ field

\begin{equation}
{\cal W}(\Psi_i ; \Pi_i) = \int D R_i e^{-i\int d^3x \Pi_i(\vec x)
R_i(\vec x)} \rho(\Psi_i+\frac{R_i}{2};\Psi_i-\frac{R_i}{2})~~;~~
\rho(\Psi_i+\frac{R_i}{2};\Psi_i-\frac{R_i}{2})= \int D \Pi_i
e^{i\int d^3x \Pi_i(\vec x) R_i(\vec x)}{\cal W}(\Psi_i ;
\Pi_i)\label{wignerrho}
\end{equation}
The boundary conditions on the $\Phi$ path integral given by
(\ref{condsfi}) translate into the following boundary conditions on
the center of mass and relative variables
\begin{equation}
\Psi(\vec x,t=0)= \Psi_i \; \; ; \; \; R(\vec x,t=0)= R_i
\label{bcwig}
\end{equation}

\noindent furthermore, the boundary condition (\ref{endbc}) yields
the following boundary condition for the relative field

\be \label{Rinfty} R(\vec{x},t=\infty)= 0. \ee

This observation will be important in the steps that follow. In
terms of the spatial Fourier transforms of the center of mass and
relative  variables (\ref{wigvars}) introduced above, integrating by
parts and accounting for the boundary conditions (\ref{bcwig}), the
non-equilibrium effective action (\ref{influfunc}) becomes:
\begin{eqnarray}
iL_{eff}[\Psi,R] & = & \int_0^{\infty} dt \sum_{\vec k} \left\{-i R_{-\vec k}\left( \ddot{\Psi}_{\vec k}(t)+(k^2+m^2)\Psi_k(t) \right)\right\} \nonumber \\
                 & - & \int_0^{\infty} dt \int_0^{\infty} dt' \left\{\frac{1}{2}R_{-\vec k}(t)R_{\vec k}(t'){\cal K}_k(t-t') + R_{-\vec k}(t)
i\Sigma^R_k(t-t') \Psi_{\vec k}(t') \right \} \nonumber \\
& + & \int d^3x R_i(\vec x) \dot{\Psi}(\vec x,t=0)
\label{efflanwig}
\end{eqnarray}
where the last term arises after the integration by parts in time,
using the boundary conditions (\ref{bcwig}) and (\ref{Rinfty}).
The kernels in the above effective Lagrangian are given by (see
eqns. (\ref{ggreat}-\ref{antitimeordered}))
\begin{eqnarray}
\mathcal{K}_k(t-t') & = &
\frac{g^2}{2} \left[{\cal G}_k^>(t-t')+{\cal G}_k^<(t-t') \right] \label{kernelkappa} \\
i\Sigma^{R}_k(t-t') & = &  g^2 \left[{\cal G}_k^>(t-t')-{\cal
G}_k^<(t-t') \right]\Theta(t-t') \equiv
i\Sigma_{k}(t-t')\Theta(t-t') \label{kernelsigma}
\end{eqnarray}

The term quadratic in the relative variable $R$ can be written in terms of a stochastic
noise as
\begin{eqnarray}
\exp\Big\{-\frac{1}{2} \int dt \int dt' R_{-\vec k}(t){\cal
K}_k(t-t')R_{\vec k}(t')\Big\} = \int {\cal D}\xi
\exp\Big\{-\frac{1}{2} \int dt \int dt' ~~ \xi_{\vec k}(t)
{\cal K}^{-1}_k(t-t')\xi_{-\vec k}(t')  \nonumber \\
 +i \int dt ~~ \xi_{-\vec k}(t) R_{\vec k}(t)\Big\}
\label{noisefunc}
\end{eqnarray}

The non-equilibrium generating functional can now be written in the
following form
\begin{eqnarray}
{\cal Z}  & = &   \int D \Psi_i \int D \Pi_i \int {\cal D} \Psi {\cal D}R {\cal D}\xi ~~
{\cal W}(\Psi_i;\Pi_i) DR_i
 e^{i \int d^3x R_i(\vec x) \left(\Pi_i (\vec x)-\Psi(\vec x,t=0)\right)}
 {\cal P}[\xi] \label{genefunc} \\
&  & \exp\left\{-i \int_0^{\infty} dt ~~ R_{-\vec k}(t) \left[
\ddot{\Psi}_{\vec k}(t)+(k^2+m^2)\Psi_{\vec k}(t)+\int dt' ~~ \Sigma^{R}_k(t-t')\Psi_{\vec k}(t')-\xi_{\vec k}(t) \right] \right\} \nonumber \\
{\cal P}[\xi] & = & \exp\left\{-\frac{1}{2} \int_0^{\infty} dt
\int_0^{\infty} dt' ~~ \xi_{\vec k}(t) {\cal K}_k^{-1}(t-t')
\xi_{-\vec k}(t') \right\} \label{probaxi}
\end{eqnarray}

The functional integral over $R_i$ can now be done, resulting in a functional delta function,
that fixes the boundary condition $\dot{\Psi}(\vec x,t=0) =
\Pi_i(\vec x)$.

Finally the path integral over the relative variable can be
performed, leading to a functional delta function and the final form
of the generating functional  given by
\begin{eqnarray}
{\cal Z}   =   \int D \Psi_i   D \Pi_i ~~{\cal W}(\Psi_i;\Pi_i)
 {\cal D} \Psi {\cal D}\xi ~~ {\cal P}[\xi]\,
 \delta\left[
\ddot{\Psi}_{\vec k}(t)+(k^2+m^2)\Psi_{\vec k}(t)+\int_0^{t} dt'
~\Sigma_k(t-t')\Psi_{\vec k}(t')-\xi_{\vec k}(t) \right]
\label{deltaprob}
\end{eqnarray}
with the boundary conditions on the path integral on $\Psi$ given by
\begin{equation}
\Psi(\vec x,t=0) = \Psi_i(\vec x) \; \; ; \; \; \dot{\Psi}(\vec
x,t=0)= \Pi_i(\vec x) \label{bcfin}
\end{equation}
\noindent where we have used the definition of $\Sigma^{R}_k(t-t')$
in terms of $\Sigma_k(t-t')$ given in equation (\ref{kernelsigma}).

The meaning of the above generating functional is the following: in
order to obtain correlation functions of the center of mass Wigner
variable $\Psi$ we must first find the solution of the classical
{\em stochastic} Langevin equation of motion
\begin{eqnarray}
&& \ddot{\Psi}_{\vec k}(t)+(k^2+m^2)\Psi_{\vec k}(t)+\int_0^t dt'
~ \Sigma_{k}(t-t')
\Psi_{\vec k}(t')=\xi_{\vec k}(t) \nonumber \\
&& \Psi_{\vec k}(t=0)= \Psi_{i,\vec k} ; ~~
\dot{\Psi}_{\vec k}(t=0)= \Pi_{i,\vec k}
\label{langevin}
\end{eqnarray}
for arbitrary noise term $\xi$ and then average the products of
$\Psi$ over the stochastic noise with the Gaussian probability
distribution ${\cal P}[\xi]$ given by (\ref{probaxi}), and finally
average over the initial configurations $\Psi_i(\vec x);
\Pi_i(\vec x)$ weighted by the Wigner function ${\cal
W}(\Psi_i,\Pi_i)$, which plays the role of an initial
 phase space distribution function.

 Calling
the solution of (\ref{langevin}) $\Psi_{\vec
k}(t;\xi;\Psi_i;\Pi_i)$,  the two point correlation function, for
example, is given by
\begin{equation}
\langle \Psi_{-\vec k}(t) \Psi_{\vec k}(t') \rangle = \int {\cal
D}[\xi] {\cal P}[\xi] \int D \Psi_i \int D\Pi_i ~~{\cal
W}(\Psi_i;\Pi_i) \Psi_{\vec k}(t;\xi;\Psi_i;\Pi_i) \Psi_{-\vec
k}(t';\xi;\Psi_i;\Pi_i) \label{expecvalcm}
\end{equation}

We note that in computing the averages and using the functional
delta function to constrain the configurations of $\Psi$ to the
solutions of the Langevin equation, there is the Jacobian of the
operator $d^2/dt^2 +(k^2+m^2)+\int dt' \Sigma^{ret}_{k}(t-t')$ which
however, is independent of the field and cancels between numerator
and denominator in the averages.

This formulation establishes the connection with a
\emph{stochastic} problem and is similar to the
Martin-Siggia-Rose\cite{MSR} path integral formulation for
stochastic phenomena. There are two different averages:

\begin{itemize}
\item{ The average over the  stochastic noise term, which up to
this order is Gaussian. We denote the average of a functional
$\mathcal{F}[\xi]$  over the noise with the probability distribution
function $P[\xi]$ given by eqn. (\ref{probaxi}) as

\begin{equation}\label{stocha}
\langle \langle \mathcal{F}[\xi] \rangle \rangle \equiv \frac{\int
\mathcal{D}\xi P[\xi] \mathcal{F}[\xi]}{\int \mathcal{D}\xi P[\xi]}.
\end{equation}

Since the noise probability distribution function is Gaussian the
only necessary correlation functions for the noise are given by

\begin{equation}
\langle \langle \xi_{\vec{k}}(t)\rangle \rangle =0 \; , \; \langle
\langle \xi_{\vec{k}}(t)\xi_{\vec{k}'}(t')\rangle \rangle =
{\mathcal K}_k(t-t')\,\delta^{3}(\vec{k}+\vec{k}')
\label{noisecorrel}
\end{equation}
\noindent and  the higher order correlation functions are obtained
from Wick's theorem. Because the noise kernel
$\mathcal{K}_k(t-t')\neq \delta(t-t')$ the noise is
\emph{colored}. }

\item{The average over the initial conditions with the Wigner
distribution function ${\cal W}(\Psi_i,\Pi_i)$ which we denote as

\begin{equation}
\overline{\mathcal{A}[\Psi_i,\Pi_i]} \equiv  \frac{\int D \Psi_i
\int D\Pi_i ~~{\cal W}(\Psi_i;\Pi_i) \mathcal{A}[\Psi_i,\Pi_i]}{\int
D \Psi_i \int D\Pi_i ~~{\cal W}(\Psi_i;\Pi_i) } \label{iniaverage}
\end{equation}
In what follows we will consider a Gaussian initial Wigner
distribution function with vanishing mean values of $\Psi_i;\Pi_i$
with the following averages:

\begin{eqnarray}
&&\overline{\Psi_{i,\vec k}\Psi_{i,-\vec k}} =
\frac{1}{2W_k}\left[1+2\mathcal{N}_{b,k}\right]\label{psi2}\\
&&\overline{\Pi_{i,\vec k}\Pi_{i,-\vec k}} =
\frac{W_k}{2}\left[1+2\mathcal{N}_{b,k}\right]\label{pi2} \\
&& \overline{\Pi_{i,\vec k}\Psi_{i,-\vec k}+\Psi_{i,\vec k}
\Pi_{i,-\vec k}} = 0 \label{psipi}
\end{eqnarray}

\noindent where $W_k$ is a reference frequency. Both $W_k$ and
$\mathcal{N}_{b,k}$ characterize the initial gaussian density
matrix. Such a density matrix describes a free field theory of
particles with frequencies $W_k$.  The averages
(\ref{psi2},\ref{pi2}) are precisely the expectation values obtained
in a free field Fock state with $\mathcal{N}_{b,k}$ number of free
field quanta of momentum $k$ and frequency $W_k$ or a free field
density matrix which is diagonal in the Fock representation of a
free field with frequency $W_k$.  This can be seem simply by writing
the field and canonical momentum in terms of the usual creation and
annihilation operators of Fock quanta of momentum $k$ and frequency
$W_k$. While this is a particular choice of initial state, we will
see below that the distribution function becomes insensitive to it
after a time scale longer than the quasiparticle relaxation time.  }

\end{itemize}

The average in the time evolved full density matrix is therefore
defined by

\begin{equation}
\langle \cdots \rangle \equiv \overline{\langle \langle
\cdots\rangle \rangle}~~. \label{totave}
\end{equation}

\subsection{Fluctuation and Dissipation:}

 From the expression (\ref{kernelsigma}) for the self-energy and the Wightmann
functions (\ref{ggreat},\ref{lesser}) which are obtained as
averages in the  equilibrium density matrix of the $\chi$ fields
(bath), we now obtain a dispersive representation for the kernels
$\mathcal{K}_k(t-t');\Sigma^{R}_k(t-t')$.
 This is achieved by explicitly writing the expectation value in
terms of energy eigenstates of the bath,  introducing the identity
in this basis, and using the time evolution of the Heisenberg field
operators to obtain
\begin{eqnarray}
g^2{\cal G}_k^>(t-t') & = &  \int^{\infty}_{-\infty} d\omega ~ \sigma^>_{\vec k}(\omega)~e^{i\omega(t-t')}  \label{specrepgreat} \\
g^2{\cal G}_k^<(t-t') & = &  \int^{\infty}_{-\infty} d\omega ~
\sigma^<_{\vec k}(\omega)~e^{i\omega(t-t')} \label{specrepless}
\end{eqnarray}

\noindent with the spectral functions

\begin{eqnarray}
\sigma^>_{\vec k}(\omega) & = &  \frac{g^2}{\mathcal{Z}_b}
\sum_{m,n}e^{-\beta E_n}
\langle n| {\cal O}_{\vec k}(0) |m \rangle \langle m| {\cal O}_{-\vec k}(0) |n \rangle \, \delta(\omega-(E_n-E_m)) \label{siggreat} \\
\sigma^<_{\vec k}(\omega) & = &  \frac{g^2}{\mathcal{Z}_b}
\sum_{m,n} e^{-\beta E_m}
 \langle n| {\cal O}_{-\vec k}(0) |m \rangle \langle m| {\cal O}_{\vec k}(0) |n
 \rangle \, \delta(\omega-(E_m-E_n))
 \label{sigless}
\end{eqnarray}

\noindent where $\mathcal{Z}_b=\mathrm{Tr}\,e^{-\beta H_{\chi}}$ is
the equilibrium partition function of the ``bath''. Upon relabelling
$m \leftrightarrow n$ in the sum in the definition (\ref{sigless})
we find the KMS relation\cite{kapusta,lebellac}

\begin{equation}
\sigma^<_{ k}(\omega)  = \sigma^>_{ k}(-\omega) = e^{\beta \omega}
\sigma^>_{ k}(\omega) \label{KMS}
\end{equation}

\noindent where we have used parity and rotational invariance in
the second line above to assume that the spectral functions only
depend of the absolute value of the momentum.

Using the spectral representation of the $\Theta(t-t')$ we find
the following representation for the retarded self-energy

\begin{equation}
\Sigma^R_k(t-t')= \int_{-\infty}^{\infty}\frac{dk_0}{2\pi}
e^{ik_0(t-t')} \widetilde{\Sigma}^R(k,k_0) \label{sigreta}
\end{equation}

\noindent with

\begin{equation}\label{sigofomega}
\widetilde{\Sigma}^R(k,k_0)=\int_{-\infty}^{\infty}d\omega
\frac{[\sigma^>_k(\omega)-\sigma^<_k(\omega)]}{\omega-k_0+i\epsilon}
\end{equation}

Using the condition (\ref{KMS}) the above spectral representation
can be written in a more useful manner as

\begin{eqnarray}
\widetilde{\Sigma}^R(k,k_0) & = &
-\frac{1}{\pi}\int_{-\infty}^{\infty}d\omega
\frac{\mathrm{Im}\widetilde{\Sigma}^R(k,\omega)}{\omega-k_0+i\epsilon}\,,\label{specsigret}\eea

\noindent where the imaginary part of the self-energy is given by

\bea \mathrm{Im}\widetilde{\Sigma}^R(k,\omega) & = & \pi
\sigma^>_k(\omega)\left[e^{\beta \omega}-1\right] \label{imagpart}
\end{eqnarray}

\noindent and is clearly positive for $\omega >0$.  Equation
(\ref{KMS}) entails that the imaginary part of the retarded
self-energy is an odd function of frequency, namely

\begin{equation}
\mathrm{Im}\widetilde{\Sigma}^R(k,\omega) = -
\mathrm{Im}\widetilde{\Sigma}^R(k,-\omega)\; .  \label{odd}
\end{equation}

The relation (\ref{imagpart}) leads to the  following  results
which will  be useful later

\begin{eqnarray}
\sigma^>_k(\omega)& = &
\frac{1}{\pi}\mathrm{Im}\widetilde{\Sigma}^R(k,\omega)\,n(\omega)
\label{sigplu}\\
\sigma^<_k(\omega)& = &
\frac{1}{\pi}\mathrm{Im}\widetilde{\Sigma}^R(k,\omega)\,\left[1+n(\omega)\right]\label{sigmin}
\end{eqnarray}

Similarly from the definitions (\ref{kernelkappa}) and
(\ref{specrepgreat},\ref{specrepless}) and the condition (\ref{KMS})
we find

\begin{eqnarray}
\mathcal{K}_k(t-t') & = &\int_{-\infty}^{\infty}\frac{dk_0}{2\pi}
e^{ik_0(t-t')} \widetilde{\mathcal{K}}(k,k_0) \label{noisefou} \\
\widetilde{\mathcal{K}}(k,k_0)& = & \pi
\sigma^>_k(k_0)\left[e^{\beta k_0}+1\right] \label{noisekernel}
\end{eqnarray}

\noindent whereupon using the condition (\ref{KMS}) leads to the
followint generalized form of the fluctuation-dissipation relation

\begin{equation}
\widetilde{\mathcal{K}}(k,k_0)=\mathrm{Im}\widetilde{\Sigma}^R(k,k_0)\coth\left[\frac{\beta
k_0}{2}\right]\label{flucdiss}
\end{equation}

Thus we see that
$\mathrm{Im}\widetilde{\Sigma}^R(k,k_0)\,;\,\widetilde{\mathcal{K}}(k,k_0)$
are odd and even functions of frequency respectively.

For further analysis below we will also need the following
representation (see eqn. (\ref{kernelsigma}))

\begin{equation}
\Sigma_k(t-t') = -i \int_{-\infty}^{\infty} e^{i\omega(t-t')}
\left[\sigma^>_k(\omega)-\sigma^<_k(\omega)\right]d\omega =
\frac{i}{\pi}\int_{-\infty}^{\infty}
e^{i\omega(t-t')}\mathrm{Im}\widetilde{\Sigma}^R(k,\omega) d\omega
\label{sig}
\end{equation}

\noindent whose Laplace transform is given by

\begin{equation}
\widetilde{\Sigma}(k,s)\equiv \int^{\infty}_0 dt e^{-st}\Sigma_k(t)=
-\frac{1}{\pi} \int^{\infty}_{-\infty}
\frac{\mathrm{Im}\widetilde{\Sigma}^R(k,\omega)}{\omega+is}d\omega
\label{laplasig}
\end{equation}

This spectral representation, combined with (\ref{specsigret})
lead to the relation

\begin{equation}
\widetilde{\Sigma}^R(k,k_0)=\widetilde{\Sigma}(k,s=ik_0+\epsilon)\label{analyt}
\end{equation}

\begin{figure}
\includegraphics[height=2.5in,width=2.5in,keepaspectratio=true]{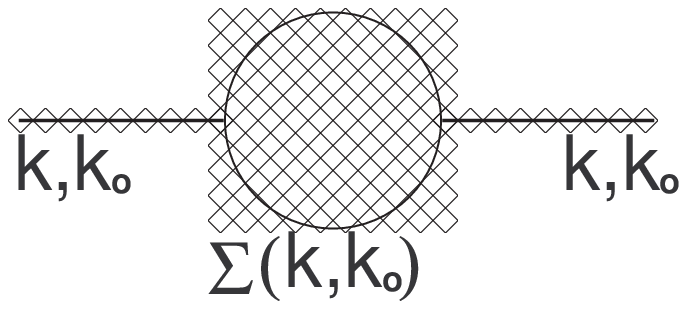}
\caption{Self-energy of $\Phi$ to lowest order in $g^2$ but to all
orders in the couplings of the fields $\chi$ amongst themselves. The
external lines correspond to the field $\Phi$.}
\label{fig:selfenergy}
\end{figure}

We highlight that the self-energy $\widetilde{\Sigma}^R(k,k_0)$ as
well as the fluctuation kernel $\widetilde{\mathcal{K}}(k,k_0)$
are \emph{to all orders} in the couplings amongst  the fields
$\chi$ but to lowest order, namely $\mathcal{O}(g^2)$ in the
coupling between the field $\Phi$ and the fields $\chi$. The
self-energy is depicted in fig.(\ref{fig:selfenergy}).

\subsection{The solution:}

The solution of the Langevin equation (\ref{langevin}) can be
found by Laplace transform. Defining the Laplace transforms

\begin{eqnarray}\label{laplapsi}
\widetilde{\Psi}_{\vec k}(s)  \equiv \int^{\infty}_0 dt
e^{-st}\Psi_{\vec k}(t)\label{laplapsi}\\
\widetilde{\xi}_{\vec k}(s)  \equiv \int^{\infty}_0 dt
e^{-st}\xi_{\vec k}(t)\label{laplapsi}
\end{eqnarray}

\noindent along with the Laplace transform of the self-energy
given by eqn. (\ref{laplasig}) we find the solution

\begin{equation}\label{solution}
\widetilde{\Psi}_{\vec k}(s)=\frac{\Pi_{i,\vec k}+s\Psi_{i,\vec
k}+\widetilde{\xi}_{\vec
k}(s)}{s^2+\omega^2_k+\widetilde{\Sigma}(k,s)}
~~;~\omega^2_k=k^2+m^2
\end{equation}
\noindent where we have used the initial conditions (\ref{bcfin}).
The solution in real time can be written in a more compact manner
as follows. Introduce the function $f_k(t)$ that obeys the
following equation of motion and initial conditions

\begin{equation}
\ddot{f}_{k}(t)+\omega^2_k\,f_{ k}(t)+\int_0^t dt' ~
\Sigma_{k}(t-t') f_{ k}(t')=0~~;~~ f_{}(t=0)= 0; ~~ \dot{f}_{
k}(t=0)=1 \label{functionf}
\end{equation}

\noindent whose Laplace transform is given by

\begin{equation}\label{laplaf}
\widetilde{f}_k(s) =
\frac{1}{s^2+\omega^2_k+\widetilde{\Sigma}(k,s)}
\end{equation}

In terms of this auxiliary function the solution of the Langevin
equation (\ref{langevin}) in real time is given by

\begin{equation}
\Psi_k(t;\Psi_i;\Pi_i;\xi) = \Psi_{i,\vec k}~ \dot{f}_k(t) +
 \Pi_{i,\vec k}~ f_k(t)+ \int^t_0
f_k(t-t')~\xi_{\vec k}(t') dt' \label{inhosolution}
\end{equation}

For the study of the number operator below we will also need the
time derivative of the solution, given by
\begin{equation}
\dot{\Psi}_k(t;\Psi_i;\Pi_i;\xi) = \Psi_{i,\vec k}~ \ddot{f}_k(t)
+
 \Pi_{i,\vec k}~ \dot{f}_k(t)+ \int^t_0
\dot{f}_k(t-t')~\xi_{\vec k}(t') dt' \label{inhosolutiondot}
\end{equation}
\noindent where we have used the initial conditions given in eqn.
(\ref{inhosolution}). From eqn. (\ref{solution}) it is clear that
the solution (\ref{inhosolution}) represents a Dyson resummation
of the perturbative expansion.

\begin{figure}[ht!]
\begin{center}
\includegraphics[height=2.5in,width=2.5in,keepaspectratio=true]{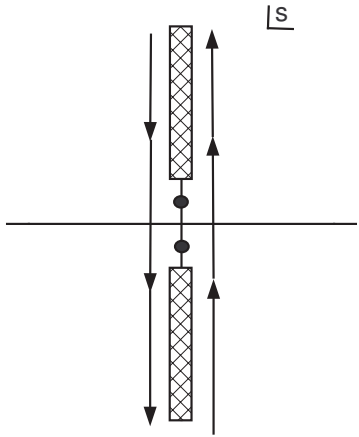}
\caption{General structure of the self-energy in the complex
s-plane. The dashed regions correspond to multiparticle cuts namely
 $\mathrm{Im}\widetilde{\Sigma}^R(k,s=i\omega+\epsilon)\neq 0$. The dots depict isolated poles. } \label{fig:cuts}
\end{center}
\end{figure}

The real time solution for $f(t)$ is found by the inverse Laplace
transform

\begin{equation}\label{bromw}
f_k(t) = \int_{C}\frac{ds}{2\pi i} \frac{e^{st}
}{s^2+\omega^2_k+\widetilde{\Sigma}(k,s)}
\end{equation}
\noindent where $C$ stands for the Bromwich contour, parallel to the
imaginary axis in the complex $s$ plane to the right of all the
singularities of $\widetilde{f}(s)$ and along the semicircle at
infinity for $\mathrm{Re}\,s < 0$. The singularities of
$\widetilde{f}(s)$ in the physical sheet are isolated single
particle poles  and multiparticle cuts along the imaginary axis.
Thus the contour can be deformed to run parallel to the imaginary
axis with a small positive real part with
$s=i\omega+\epsilon\,;\,-\infty \leq \omega \leq \infty$ , returning
parallel to the imaginary axis with
$s=i\omega-\epsilon\,;\,\infty>\omega>-\infty$, with $\epsilon =
0^+$ as depicted in fig. (\ref{fig:cuts}).

From the spectral representations (\ref{imagpart},\ref{laplasig}))
one finds that
$\widetilde{\Sigma}(k,s=i\omega\pm\epsilon)=\mathrm{Re}\widetilde{\Sigma}^R(k,\omega)\pm
\mathrm{Im}\widetilde{\Sigma}^R(k,\omega)$ and using that
$\mathrm{Im}{\widetilde{\Sigma}}^R(k,\omega)=-\mathrm{Im}\widetilde{\Sigma}^R(k,-\omega)$
we find the following solution in real time

\be \label{frho} f_k(t)= \int_{-\infty}^{\infty} \sin(\omega t) \,
\rho(k,\omega;T)\,d\omega \,, \ee

\noindent where we have introduced the spectral density

\be\label{rho} \rho(k,\omega;T) = \frac{1}{\pi}\,
\frac{~\left[\mathrm{Im}\widetilde{\Sigma}^R(k,\omega;T)+2\omega\epsilon\right]}{\left[\omega^2-\omega^2_k-\mathrm{Re}\widetilde{\Sigma}^R(k,\omega;T)
\right]^2+\left[\mathrm{Im}\widetilde{\Sigma}^R(k,\omega;T)+2\omega\epsilon\right]^2}\;,
\ee

\noindent and we have made explicit the temperature dependence of
the self-energy.

We have kept the infinitesimal $2\omega\epsilon$ with $\epsilon
\rightarrow 0^+$  since if there are isolated single particle
poles away from the multiparticle cuts for which
$\mathrm{Im}\widetilde{\Sigma}^R(k,s)= 0$  then this term ensures
that the isolated pole contribution is accounted for, namely

\be \label{SPpole}\frac{1}{\pi} ~\frac{
2\omega\epsilon}{\left[\omega^2-\omega^2_k-\mathrm{Re}\widetilde{\Sigma}^R(k,\omega)
\right]^2+\left[2\omega\epsilon\right]^2} = \mathrm{sign}(\omega)\,
\delta\left[\omega^2-\omega^2_k-\mathrm{Re}\widetilde{\Sigma}^R(k,\omega)
\right]\;.\ee

The initial condition $\dot{f}_k(t=0)=1$ leads to  the following
sum rule
\begin{equation}
\int_{-\infty}^{\infty} \frac{d\omega}{\pi} ~\frac{\omega
\left[\mathrm{Im}\widetilde{\Sigma}^R(k,\omega)+2\omega\epsilon\right]}{\left[\omega^2-\omega^2_k-\mathrm{Re}\widetilde{\Sigma}^R(k,\omega)
\right]^2+\left[\mathrm{Im}\widetilde{\Sigma}^R(k,\omega)+2\omega\epsilon\right]^2}
=1 \label{sumrule}
\end{equation}

\section{Counting particles: the number operator}\label{sec:count}

In an interacting theory the definition of a particle number
requires careful consideration. To begin with, a distinction must
be made  between physical particles that appear in asymptotic
states and can be counted by a detector, from unstable particles
or resonances which have  a finite lifetime and  decay into other
particles. Resonances are not asymptotic states,  do not
correspond to eigenstates of a Hamiltonian and their presence is
inferred from virtual contributions to cross sections. In an
interacting theory virtual processes turn a bare particle into a
physical particle by dressing the bare particle with a cloud of
virtual excitations. Physical particles correspond to asymptotic
states and are eigenstates of the full (interacting) Hamiltonian
with the physical mass. These physical particles correspond to
\emph{real} poles in the Green's functions or propagators in the
complex frequency plane. In the exact vacuum state, the propagator
of the field associated with the physical particles features poles
below the multiparticle continuum at the exact frequencies and
with a residue given by the wave function renormalization constant
$Z$. The wave function renormalization determines the overlap
between the bare and interacting single particle states. Lorentz
invariance of the vacuum state entails that the exact frequencies
are of the form $\Omega_k = \sqrt{k^2+m^2_P}$, where $m_P$ is the
physical mass and that the wave function renormalization is
independent of the momentum $k$. In asymptotic theory, the spatial
Fourier transform of the field operator $\hat{\Phi}_{\vk}(t)$
obeys the (weak) asymptotic condition

\be\label{asycond} \hat{\Phi}_{\vk}(t)|0\rangle~~
\overrightarrow{t\rightarrow \infty}~~\sqrt{
\frac{Z}{{2\Omega_k}}}\,e^{i\Omega_k\,t}\,a^{\dagger}_{out}|0\rangle
\equiv \sqrt{
\frac{Z}{{2\Omega_k}}}\,e^{i\Omega_k\,t}\,|1_{\vk}\rangle\, ,\ee

\noindent where $|1_{\vk}\rangle$ is the state with one physical
particle.

In a medium at finite temperature there are no asymptotic states,
each particle, even when stable in vacuum acquires a width in a
medium either by collisional  processes (collisional broadening)
or other processes such as Landau damping. The width acquired by a
physical particle in a medium is a  consequence of the interaction
between the physical particle and  the  excitations in the medium.
 In particular the medium-induced width is necessary to ensure that physical particles relax to a state of
thermal equilibrium with  the medium. The relaxation rate is a
measure of the width of the particle in the medium. Therefore in a
medium a physical particle becomes a \emph{quasiparticle} with a
medium modification of the dispersion relation and a width.

Thus the question  arises as to what particles are ``counted'' by
a definition of a  distribution function, namely, a decision must
be made  to count either physical particles or
\emph{quasiparticles}.

 One can envisage counting physical particles by introducing a detector in
 the medium. Such detector must be calibrated so as to ``click''
 every time it finds a particle with given characteristics. A
 detector that has been calibrated to measure physical particles
 in a scattering experiment for example, will measure the energy
 and the momentum (and any other good quantum numbers) of a
 particle. Every time that the detector measures a momentum $\vk$
 and an energy $\Omega_k$  determined by the dispersion relation of
 the physical particle (as well as other available quantum numbers), it counts this ``hit'' as one particle.

  Once this detector has been calibrated
 in this manner, for example by carrying out a scattering
 experiment in the vacuum, we can insert this detector in a medium
 and let it count the physical particles \emph{in the medium}.

Counting \emph{quasiparticles} entails a different calibration of
the detector which must account for the properties of the medium in
the definition of a quasiparticle. The first obstacle in such
calibration is the fact that a quasiparticle does not have a
definite dispersion relation because its spectral density features a
width, namely a quasiparticle is not associated with a sharp energy
but with a continuum  distribution of energies. How much of this
distribution will be accepted by the detector in its definition of a
quasiparticle, will depend on the filtering process involved in
accepting a quasiparticle, and so cannot be unique. Therefore
statements about measuring a distribution of quasiparticles are
somewhat ambiguous.

In this article we focus on the first strategy, by counting only
\emph{physical particles}. Hence we \emph{propose} a number operator
that ``counts'' the physical particle states of mass $m_P$ that a
detector will measure for example in a scattering experiment at
asymptotically long times. Asymptotic theory and the usual reduction
formula suggest the following \emph{definition} of an interpolating
number operator that counts the number of physical (stable)
particles in a state

\begin{equation}
\hat{N}_k(t) = \frac{1}{2\Omega_k\,Z} \left \{
\hat{\dot{\Phi}}_{\vec k}(t) \hat{\dot{\Phi}}_{-\vec k}(t)+
\Omega^2_k \hat{{\Phi}}_{\vec k}(t) \hat{{\Phi}}_{-\vec k}(t)
\right\} -\mathcal{C}_k \label{numberop}
\end{equation}
\noindent where $Z$ is  the wave function renormalization, namely
the residue of the single (physical) particle pole in the exact
propagator, $\Omega_k=\sqrt{k^2+m^2_P}$ is the renormalized physical
frequency and the normal ordering constant $\mathcal{C}_k$ will be
adjusted so as to include renormalization effects. In free field
theory $\Omega_k=\omega_k
=\sqrt{k^2+m^2}\,,Z=1\,,\,\mathcal{C}_k=1/2$. However, in asymptotic
theory the field $\Phi$ creates a single particle state of momentum
$k$ and mass $m_P$ with amplitude $\sqrt{Z}$ out of the exact
vacuum.

The quantity $\mathcal{C}_k$ arises from the necessity of
redefining the normal ordering  for the correct identification of
the particle number in an interacting field theory. It will be
fixed below by requiring that the expectation value of
$\hat{N}_k(t)$ vanishes in the exact vacuum state at
asymptotically long time. Alternatively this constant can be
extracted from the equal time limit of the operator product
expansion.

The approach that we follow is to consider an initial factorized
density matrix corresponding to a tensor product of a density matrix
of the field $\Phi$ and a thermal bath of the fields $\chi$.  This
initial state will evolve in time with the full interacting
Hamiltonian, leading to transient phenomena which results in the
dressing of the bare particles by the virtual excitations. At
asymptotically long times the bare particle is fully dressed into
the physical particle, and at finite temperature, a quasiparticle.
The time evolution of the interpolating number operator will reflect
this transient stage and the dynamics of the dressing of the bare
into the physical state.  Since the thermal bath is stationary, the
distribution of physical particles in the bath will be extracted
from the asymptotic long time limit of the expectation value of the
interpolating Heisenberg number operator $\hat{N}_k(t)$ in the
initial state.

The expectation value of $\hat{N}_k(t)$ is related to the real-time
correlation functions of the field $\Phi$ as follows
\begin{equation}
\langle \hat{N}_k(t) \rangle = \frac{1}{4\Omega_k\,Z}
\left(\frac{\partial}{\partial t} \frac{\partial}{\partial t'}+
\Omega^2_k\right) \Bigg[g^>_k(t,t')+g^<_k(t,t')\Bigg]_{t=t'}
-\mathcal{C}_k \label{expnumb}
\end{equation}
where the non-equilibrium correlation functions are given by
\begin{eqnarray}
&& \langle \Phi^+_{\vec k}(t) \Phi^+_{-\vec k}(t') \rangle =
g^>_k(t,t') \Theta(t-t')+g^<_k(t,t')\Theta(t'-t) \label{timeord}\\
&& \langle \Phi^-_{\vec k}(t) \Phi^-_{-\vec k}(t') \rangle =
g^>_k(t,t') \Theta(t'-t)+g^<_k(t,t')\Theta(t-t') \label{antitimeord}\\
 &&\langle \Phi^-_{\vec k}(t) \Phi^+_{-\vec k}(t') \rangle  =  g^>_k(t,t')
\label{phigreat} \\
 &&\langle \Phi^-_{-\vec k}(t') \Phi^+_{\vec k}(t)
\rangle  =  g^<_k(t,t') \label{philess}
\end{eqnarray}

The symmetrized definition of the expectation value (\ref{expnumb})
has been chosen for convenience because, as it is shown in detail
below,  it is related to a simple correlation function of the center
of mass field $\Psi_k(t)$ given by eqn. (\ref{wigvars}). However we
could have chosen any other definition of the equal time correlation
function, such as $g^>_k(t,t); g^<_k(t,t)$ or any combination
thereof that does not involve a Heaviside step function in time for
which the time derivatives will yield spurious delta functions. It
is a straightforward exercise with the density matrix to show that
all of these alternative definitions are equivalent in the equal
time limit, since these do not involve discontinuous functions of
time.


In terms of the center of mass field $\Psi_k(t)= (\Phi^+_{\vec
k}(t)+\Phi^-_{\vec k}(t))/2$ introduced above it is straightforward
to  find that the correlation function in the bracket in
(\ref{expnumb}) is given by
\begin{equation}
\langle \Psi_{\vec k}(t) \Psi_{-\vec k}(t') \rangle = \frac{1}{2} \left[g^>_k(t,t')+g^<_k(t,t') \right] \label{CMcorre}
\end{equation}
and the occupation number can be written in terms of the center of
mass Wigner variable  introduced in eqn. (\ref{wigvars}) as follows
\begin{equation}
\langle \hat{N}_k(t) \rangle = \frac{1}{2\Omega_k Z}\left[ \langle
\dot{\Psi}_{\vec k}(t) \dot{\Psi}_{-\vec k}(t)\rangle +\Omega^2_k
\langle \Psi_{\vec k}(t) \Psi_{-\vec
k}(t)\rangle\right]-\mathcal{C}_k \label{CMnumber}
\end{equation}

\noindent where the expectation values are obtained as in eqn.
(\ref{totave}) and $ \Psi_{\vec k}(t)$ is the solution of the
Langevin equation given by
(\ref{inhosolution},\ref{inhosolutiondot}).

A straightforward calculation implementing eqn.  (\ref{totave})
writing the noise in terms of its temporal Fourier transform and
using the Fourier representation of the noise kernel
(\ref{noisefou}) leads to the following result

\begin{eqnarray}
N_k(t) \equiv \langle \hat{N}_k(t) \rangle & = &
\frac{1}{2\Omega_k Z}\left\{
\frac{1}{2W_k}\left[1+2\mathcal{N}_{b,k}\right]\left[
\ddot{f}^2_k(t) + ( \Omega^2_k+W^2_k) ~ \dot{f}^2_k(t) +
\Omega^2_k W^2_k~ f^2_k(t) \right] \right. \nonumber \\
&  & \left. + \int_{-\infty}^{\infty} \frac{d\omega}{2\pi}
\widetilde{\cal K}(k,\omega) \left[ |\mathcal{F}_k(\omega,t)|^2 +
\Omega^2_k |\mathcal{H}_k(\omega,t)|^2 \right] \right\}
-\mathcal{C}_k \label{finumber}
\end{eqnarray}

\noindent where we have introduced

\begin{eqnarray}
&& \mathcal{H}_k(\omega,t)= \int_0^t d\tau {f}_k(\tau)e^{-i\omega
\tau}\label{funH}\\
 && \mathcal{F}_k(\omega,t)= \int_0^t d\tau
\dot{f}_k(\tau)e^{-i\omega \tau}\label{funF}
\end{eqnarray}

\noindent $f_k(t)$ is given in eqn. (\ref{frho}) and the
fluctuation kernel  $\widetilde{\cal K}(k,\omega) $ is given by
eqn. (\ref{flucdiss}).

The result (\ref{finumber})  for the time evolution of the
distribution function, along with the expressions
(\ref{funH},\ref{funF}) clearly highlights the
\emph{non-Markovian} nature of the evolution. The integrals in
time in (\ref{funH},\ref{funF}) include \emph{memory} of the past
evolution. This is one of the most important aspects that
distinguishes the quantum kinetic approach from the usual
Boltzmann equation. We will contrast these aspects in section
(\ref{sec:boltzkin}).

\subsection{Counting physical particles in a thermal bath.}

\begin{figure}[ht!]
\includegraphics[height=2in,width=2in,keepaspectratio=true]{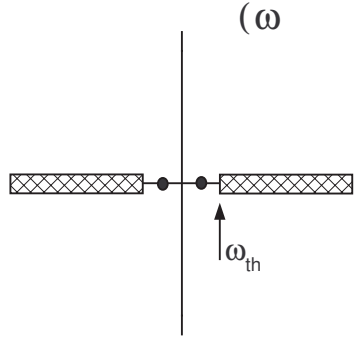}
\caption{Spectral density $\rho_k(\omega,T=0)$ for stable
particles. The dots represent the isolated poles at $\pm \Omega_k$
and the shaded regions the multiparticle cuts. $\omega_{th}$ is
the lowest multiparticle threshold.  } \label{fig:stable}
\end{figure}

In the vacuum the spectral density of the field $\Phi$ which
describes a physical particle is depicted in fig.
(\ref{fig:stable}). It features isolated poles along the real axis
in the physical sheet in the complex frequency ($\omega$)
 plane   at the position of the \emph{exact} single particle dispersion relation
$\Omega_k$ with $|\Omega_k|<|\omega_{th}|$ where $\omega_{th}$ is
the lowest multiparticle threshold.

As mentioned above, in a medium stable physical particles acquire a
width  as a consequence of the interactions with physical
excitations, and   become \emph{quasiparticles}. The width can
originate in several different processes such as collisions or
Landau damping. The poles  move off the physical sheet into the
second (or higher) Riemann sheet in the complex $\omega$ plane, thus
becoming a resonance. This is the statement that there are no
asymptotic states in the medium.

The analytic structure of the spectral density at finite temperature
is in general fairly complicated. While at zero temperature the
multiparticle thresholds are above the light cone $|\omega| > k$,
at finite temperature (or density) there appear branch cuts with
support below the light cone\cite{kapusta,lebellac,weldon,brapis}.
However a general statement in a medium is that the poles associated
with stable particles in vacuum (along the real axis in the physical
sheet) move off the physical sheet and the spectral density does not
feature isolated poles but only branch cut singularities in the
physical sheet, associated with multiparticle processes in the
medium.

In perturbation theory  the resonance is very close to the real
axis (but in the second or higher Riemann sheet) and the width is
very small as compared with the position of the resonance. We will
study a particular example in the next section.

In perturbation theory the spectral density $\rho(k,\omega,T)$
(\ref{rho}) features a sharp peak at the position of the
\emph{quasiparticle} ``pole'' which is determined by

\be \label{QPole}
\mathcal{W}^2_k(T)-\omega^2_k-\mathrm{Re}\widetilde{\Sigma}^R(k,\mathcal{W}_k(T);T)=0
\ee

 Near the quasiparticle ``poles'' the spectral
density is well described by the Breit-Wigner approximation

\be \label{rhoBW} \rho_{BW}(k,\omega;T) \simeq
\frac{\mathcal{Z}_k(T)}{2\mathcal{W}_k(T)}~\frac{1}{\pi}
 \frac{\mathrm{sign}(\omega)\,\Gamma_k(T)}{(|\omega|-\mathcal{W}_k(T))^2+\Gamma^2_k(T)}\;,  \ee

\noindent where $\mathcal{W}_k(T)$ is determined by eqn.
(\ref{QPole}) and the finite temperature residue and width are
given by

\be \label{ZT} \frac{1}{\mathcal{Z}_k(T)} = \Bigg[
1-\frac{1}{2\mathcal{W}_k(T)} \frac{\partial
\mathrm{Re}\widetilde{\Sigma}^R(k,\omega;T)}{\partial
\omega}\Bigg]_{\omega=\mathcal{W}_k(T)} \ee

\be\label{GammaT} \Gamma_k(T) =  \mathcal{Z}_k
\frac{\mathrm{Im}\widetilde{\Sigma}^R(k,\mathcal{W}_k(T);T)}{2
\mathcal{W}_k(T)} \ee

At zero temperature of the bath, the (quasi) particle dispersion
relation $\mathcal{W}_k(T)$ is identified with the dispersion
relation of the stable physical particle, namely the ``on-shell''
pole, the residue $\mathcal{Z}_k(T)$ is identified with the
wavefunction renormalization constant $Z$  which is the residue at
the on-shell pole for the physical particle, and the width
vanishes at zero temperature since the particle is stable in the
vacuum, namely

\bea \mathcal{W}_k(T=0) & = &  \Omega_k
\label{T0pole}\\\mathcal{Z}_k(T=0)  & = &  Z \label{T0Z}\\
\Gamma_k(T=0) & = & 0 \label{T0G}\eea

In the Breit-Wigner  approximation the real time solution is
easily found to be

\be\label{RTTf} f^{BW}_k(t) \simeq \mathcal{Z}_k(T)
\frac{\sin\left[\mathcal{W}_k(T)\,t\right]}{\mathcal{W}_k(T)}\,
e^{-\Gamma_k(T)\,t} \ee

This solution describes the relaxation of single
\emph{quasiparticles}, where $\mathcal{W}_k(t)$ is the quasiparticle
dispersion relation  and $\Gamma_k(T)$ is the quasiparticle decay
rate.

The asymptotic long time limit of the distribution function
(\ref{finumber}) is obtained by using the following identities

\begin{eqnarray}
{\cal H}_k(\omega,\infty) = \int_0^{\infty}
e^{-i(\omega-i\epsilon)t}
f_k(t) dt & = & \widetilde{f}_k(s=i\omega+\epsilon) \label{hinfty} \\
 {\cal F}_k(\omega,\infty) = \int_0^{\infty} e^{-i(\omega-i\epsilon)t}
\dot{f}_k(t) dt & = & i\omega \widetilde{f}_k(s=i\omega+\epsilon)
\label{finfty}
\end{eqnarray}
where $\widetilde{f}_k(s)$ is the Laplace transform of $f_k(t)$
given by eqn. (\ref{laplaf})  and in (\ref{finfty}) we have
integrated by parts, used the initial condition $f_k(0)=0$ and
introduced a convergence factor $\epsilon \rightarrow 0^+$. Hence
the expectation value of the interpolating number operator in the
asymptotic long-time limit is given by

\be \label{Ninfty} N_k(\infty) =\int_{0}^{\infty}
\Bigg(\frac{\omega^2+\Omega^2_k}{2\,Z\,\Omega_k}\Bigg)\left[1+2n(\omega)\right]\rho(k,\omega,T)\,{d\omega}\,
-\mathcal{C}_k \;,\ee

\noindent where $n(\omega)$ is the Bose-Einstein distribution
function and we have used the fluctuation-dissipation relation
(\ref{flucdiss})  as well as eqn. (\ref{laplaf}) which lead to the
$\rho(k,\omega,T)$ in (\ref{Ninfty}). The dependence of the
asymptotic distribution function on the spectral density is a
consequence of the fluctuation-dissipation relation
(\ref{flucdiss}) as well as the \emph{non-Markovian} time
evolution as displayed in (\ref{hinfty},\ref{finfty}).

The real time solution (\ref{RTTf}) clearly reveals that the
asymptotic limit is reached for $t > \tau_k = 1/2\Gamma_k(T)$
where $\Gamma_k(T)$ is the quasiparticle relaxation rate. The
distribution function at $t >> \tau_k$ does not depend on the
initial distribution $\mathcal{N}_{b,k}$ or the reference
frequencies $W_k$. Therefore at times longer than the
quasiparticle relaxation time the distribution function becomes
\emph{independent of the initial conditions}. This is to be
expected if the state reaches thermal equilibrium with the bath,
since in thermal equilibrium there is no memory of the initial
conditions or correlations.

The integral term in the asymptotic distribution (\ref{Ninfty}) is
easily understood as \emph{full thermalization} from the following
argument.

Let us consider the correlations functions $g^>_k(t,t');
g^<_k(t,t') $ given by eqns. (\ref{phigreat},\ref{philess}). In
\emph{thermal equilibrium} they have the spectral representation

\bea g^>_k(t,t')  & = &  \int \rho^>(k,\omega;T)\,
e^{i\omega(t-t')}
d\omega \label{specgreat} \\
g^<_k(t,t')  & = &  \int \rho^<(k,\omega;T)\, e^{i\omega(t-t')}
d\omega \label{specless} \eea

\noindent where

\bea \rho^>(k,\omega;T) & = &  \frac{1}{\mathcal{Z}_T}
\sum_{m,n}e^{-\beta E_n}
\langle n| {\Phi}_{\vec k}(0) |m \rangle \langle m| {\Phi}_{-\vec k}(0) |n \rangle \, \delta(\omega-(E_n-E_m)) \label{rogreat} \\
\rho^<(k,\omega;T) & = &  \frac{1}{\mathcal{Z}_T} \sum_{m,n}
e^{-\beta E_m}
 \langle n| {\Phi}_{-\vec k}(0) |m \rangle \langle m| {\Phi}_{\vec k}(0) |n
 \rangle \, \delta(\omega-(E_m-E_n))\,.
 \label{roless}\eea

Where $\mathcal{Z}_T$ is the thermal equilibrium partition
function. A straightforward re-labelling of indices leads to the
relation

\be \rho^<(k,\omega;T) = \rho^>(k,-\omega;T)= e^{\beta
\omega}\rho^>(k,\omega;T) \ee

The spectral density is given by

\be \rho(k,\omega;T) = \rho^<(k,\omega;T)-\rho^>(k,\omega;T) \ee

\noindent leading to the relations

\bea \rho^>(k,\omega;T) & = & \rho(k,\omega;T) \,n(\omega) \\
\rho^<(k,\omega;T) & = & \rho(k,\omega;T)
\,\left[1+n(\omega)\right]\, , \eea

\noindent where $n(\omega)= 1/[e^{\beta \omega}-1]$.

Therefore \emph{in thermal equilibrium} the expectation value of
the operator term in eqns. (\ref{numberop}, \ref{expnumb}) is
given by

\be \frac{1}{4\Omega_k\,Z} \left(\frac{\partial}{\partial t}
\frac{\partial}{\partial t'}+ \Omega^2_k\right)
\Bigg[g^>_k(t,t')+g^<_k(t,t')\Bigg]_{t=t'} =
\int_{-\infty}^{\infty}
\Bigg(\frac{\omega^2+\Omega^2_k}{4\,Z\,\Omega_k}\Bigg)\left[1+2n(\omega)\right]\,\rho(k,\omega;T)\,{d\omega}\,,
\label{nope}\ee

\noindent which is precisely the integral term in the asymptotic
limit given by eqn. (\ref{Ninfty}). Therefore the expression
(\ref{Ninfty}) indicates that the excitations of the field $\Phi$
have reached a state of thermal equilibrium with the bath. The
normal ordering constant $\mathcal{C}_k$ in (\ref{Ninfty}) is a
subtraction necessary to redefine normal ordering in the
interacting theory and is defined from the operator product
expansion to yield vanishing number of particles in the vacuum.

While the asymptotic long time limit can be obtained directly from
the spectral representation of the interpolating number operator in
the equilibrium state, the real time formulation in terms of the
non-equilibrium effective action has two advantages: i) it makes
explicit the connection with the fluctuation dissipation relation
and clearly states that the equilibrium abundance is determined by
the \emph{noise} correlation function of the bath, ii) the real time
dynamics clearly shows thermalization on time scales  $t> \tau_k$.
These  statements would not be immediately recognized from the
equilibrium spectral representation.

The result (\ref{Ninfty}) becomes more illuminating in the
\emph{narrow width approximation} where the Breit-Wigner
approximation for the spectral density (\ref{rhoBW})  is
supplemented with  the narrow width limit $\Gamma_k(T) \rightarrow
0$ which leads to

\be \rho(k,\omega;T) \simeq
\frac{\mathcal{Z}_k(T)}{2\mathcal{W}_k(T)}~
 \mathrm{sign}(\omega) \delta \left(|\omega|-\mathcal{W}_k(T)\right)\;,
 \ee
 \noindent which in turn leads to the \emph{approximate} result

 \be N_k(\infty) \sim  \frac{\mathcal{Z}_k(T)}{Z}
\Bigg(\frac{\mathcal{W}^2_k(T)+\Omega^2_k}{2\,\mathcal{W}_k(T)\,\Omega_k}\Bigg)\left[\frac{1}{2}+n(\mathcal{W}_k(T))\right]
-\mathcal{C}_k  \ee

Obviously the  zero temperature pole $\Omega_k$ and residue $Z$ and
their finite temperature counterparts $\mathcal{W}_k(T),
\mathcal{Z}_k(T)$ differ by terms that are of order $g^2$, namely
perturbatively small, therefore in the narrow width approximation,
which itself is a result of the weak coupling assumption one could
write

\be N_k(\infty) \sim n(\Omega_k) +\left[\frac{1}{2} +
\mathcal{O}(g^2)-\mathcal{C}_k\right] \ee

Thus choosing the normal ordering factor $\mathcal{C}_k =1/2 +
\mathcal{O}(g^2)$ would lead to the conclusion that the physical
particles are distributed in the thermal bath with a Bose-Einstein
distribution function with the argument being the physical pole
frequency (at zero temperature). Furthermore the normal ordering
constant $\mathcal{C}_k \sim 1/2 $ is identified with the usual
normal ordering of the number operator in the free field vacuum.

In order to understand in detail the perturbative correction we
have to first decide on what are $\Omega_k,Z,\mathcal{C}_k$. The
importance of the perturbative corrections cannot be
underestimated, if the temperature of the bath is much smaller
than $\Omega_k$ the distribution function $n(\Omega_k) \ll 1$ and
the perturbative corrections can be of the same order or larger.
What should be clear from the above discussion is that in order to
make precise the perturbative correction to the abundance, we must
specify unambiguously what is being counted.

\subsubsection{Physical particles in the vacuum} The next step is to
define $\Omega_k,Z,\mathcal{C}_k$. As it was emphasized above, the
 number operator that we seek counts physical particles.
These are stable excitations off the full vacuum state of the theory
and are associated with isolated single particle poles in the
spectral density \emph{at zero temperature}.

The zero temperature limit of the asymptotic distribution function
(\ref{Ninfty}) is

\be N_k(\infty;T=0) =\int_{0}^{\infty}
\Bigg(\frac{\omega^2+\Omega^2_k}{2\,Z\,\Omega_k}\Bigg)\,\rho(k,\omega,T=0)\,{d\omega}
-\mathcal{C}_k \;,\ee

At $T=0$ the spectral density features the isolated single particle
poles away from the multiparticle continuum as depicted in fig.
(\ref{fig:stable}). The contribution from the single particle poles
to the zero temperature spectral density is given by eqn.
(\ref{SPpole}), therefore we write

\be\label{rhT0} \rho(k,\omega,T=0) =
\mathrm{sign}(\omega)\,\frac{Z}{2\Omega_k}\,\delta(\omega-\Omega_k)+\rho_{c}(k,\omega,T=0)\,,
\ee

\noindent where $\rho_c(k,\omega,T=0)$ is the continuum contribution
with support for $|\omega|>\omega_{th}$,  where $\omega_{th}$ is the
lowest multiparticle threshold, and the position of the isolated
pole satisfies

\be \label{PoleT0} \Omega^2_k - \omega^2_k -
\mathrm{Re}\widetilde{\Sigma}^R(k,\Omega_k )=0 \ee

At zero temperature Lorentz covariance implies that $\Omega^2_k =
k^2+m^2_P$, where $m_P$ is the pole mass of the physical excitations
(asymptotic states).

The residue at the single (physical) particle pole, $Z$, is given by

\be \label{ZT01} \frac{1}{Z} = \Bigg[ 1-\frac{1}{2\Omega_k}
\frac{\partial \mathrm{Re}\widetilde{\Sigma}^R(k,\omega;T)}{\partial
\omega}\Bigg]_{\omega=\Omega_k} \, .\ee

\noindent  Introducing the zero temperature form of the spectral
density (\ref{rhT0}) in the sum rule (\ref{sumrule}) the following
alternative expression is obtained.

\be \label{ZT02} Z = 1- 2\int_{\omega_{th}}^\infty
\omega\,\rho_c(k,\omega,T=0)\,d\omega \ee

Therefore the asymptotic distribution of particles in the vacuum
is given by

 \be N_k(\infty;T=0) =\frac{1}{2} +\int_{0}^{\infty}
\Bigg(\frac{\omega^2+\Omega^2_k}{2\,Z\,\Omega_k}\Bigg)\,\rho_c(k,\omega,T=0)\,{d\omega}
-\mathcal{C}_k \;,\ee

The normal ordering term $\mathcal{C}_k$ is now fixed by requiring
that for $T=0$ the vacuum state has vanishing number of physical
excitations. In other words, by requiring $N_k(\infty,T=0)=0$ we are
led to

\be\label{Cnor} \mathcal{C}_k = \frac{1}{2} + \int_{0}^{\infty}
\Bigg(\frac{\omega^2+\Omega^2_k}{2\,Z\,\Omega_k}\Bigg)\,\rho_c(k,\omega,T=0)\,{d\omega}.\ee

\noindent We have kept the lower limit in the integral to be
$\omega=0$ for further convenience, however $\rho_c(k,\omega,T=0)$
\emph{vanishes} for $|\omega|<\omega_{th}$.

Equations (\ref{PoleT0}), (\ref{ZT01},\ref{ZT02}) and (\ref{Cnor})
determine all of the parameters $\Omega_k,Z,\mathcal{C}_k$ for the
proper definition of the distribution function for physical
particles.

Hence the distribution function of \emph{physical} excitations in
equilibrium with the bath at finite temperature is finally given by
the simple expression

\be \label{Nfinal} \mathcal{N}(k,T)\equiv N_k(\infty)
=\int_{0}^{\infty}
\Bigg(\frac{\omega^2+\Omega^2_k}{2\,Z\,\Omega_k}\Bigg)\Bigg\{\left[1+2n(\omega)\right]\rho(k,\omega,T)-\rho_c(k,\omega,T=0)\Bigg\}\,{d\omega}\,
-\frac{1}{2} \;,\ee

This is the final form of the  asymptotic distribution function of
physical particles in equilibrium in the thermal bath with
$\Omega_k=\sqrt{k^2+m^2_P};Z;\mathcal{C}_k$ given by equations
(\ref{PoleT0}),(\ref{ZT01}) (or (\ref{ZT02}),(\ref{Cnor}))
respectively.

\subsection{Renormalization:}\label{renormalization}

In renormalizable theories the wavefunction renormalization
constant $Z$ is ultraviolet divergent and the expression for the
asymptotic distribution function (\ref{Nfinal}) seems to be
ambiguous. However proper renormalization as described below shows
that the asymptotic abundance is finite.

In general the imaginary part of the self-energy can be written as
a sum of a zero temperature and a finite temperature contribution,
the latter vanishing at zero temperature, thus we write

\be\label{imsigsplit}
\mathrm{Im}\widetilde{\Sigma}^R(\omega,k;T)=\mathrm{Im}\widetilde{\Sigma}^R_0(\omega,k)+\mathrm{Im}\widetilde{\Sigma}^R_T(\omega,k)\ee

Therefore the real part of the self-energy, which is obtained from
the imaginary part by a dispersion relation (Kramers-Kronig) can
also be written as  a sum of a zero temperature plus a finite
temperature contribution,

\be \label{resigsplit} \mathrm{Re}\widetilde{\Sigma}^R(\omega,k;T)
= -\frac{1}{\pi} \mathcal{P} \int^{\infty}_{0} 2k_0
\frac{\mathrm{Im}\widetilde{\Sigma}^R(k_0,k;T)}{k^2_0-\omega^2}~dk_0\equiv
\mathrm{Re}\widetilde{\Sigma}^R_0(\omega,k)+
\mathrm{Re}\widetilde{\Sigma}^R_T(\omega,k)\ee

\noindent where $\mathcal{P}$ stands for the principal part of the
integral, and we have used the fact that
$\mathrm{Im}\widetilde{\Sigma}^R(k_0,k;T)$ is an odd function of
$k_0$. Both $\mathrm{Im}\widetilde{\Sigma}^R_T(\omega,k)$ and
$\mathrm{Re}\widetilde{\Sigma}^R_T(\omega,k)$ vanish at $T=0$.

The position of the physical pole is obtained at zero temperature
from the relation (\ref{PoleT0}),

\be \label{PoleT02} \Omega^2_k - \omega^2_k -
\mathrm{Re}\widetilde{\Sigma}^R_0(k,\Omega_k )=0 \ee

The subtracted real part of the self energy is

\be\label{sub00}
\mathrm{Re}\widetilde{\Sigma}^R_0(k,\omega)-\mathrm{Re}\widetilde{\Sigma}^R_0(k,\Omega_k
 ) = \Big[1-Z^{-1}[k,\omega]\Big]
(\omega^2-\Omega^2_{k})\ee

\noindent where

\be\label{ZQ00} Z^{-1}[k,\omega] =1 + \frac{1}{\pi} \mathcal{P}
\int_{0}^{\infty} 2 k_0
\frac{\mathrm{Im}\widetilde{\Sigma}^R_0(k_0,k)}{(k_0^2-\omega^2)(k^2_0-\Omega^2_k)}~dk_0.\ee


This expression for $Z^{-1}[k,\omega]$ follows from the zero
temperature limit of the dispersive representation in eqn.
(\ref{resigsplit}).

The function $Z[k,\omega=\Omega_k]$, namely evaluated on the single
particle mass shell, is identified with the wave function
renormalization, or residue at the single particle pole at zero
temperature.


As mentioned above, in renormalizable theories $Z[k,\omega]$ is
ultraviolet logarithmically divergent, therefore it is convenient to
perform yet another subtraction of the integral term in (\ref{ZQ00})
as follows,

\be \label{Zsub} Z^{-1}[k,\omega] =   Z^{-1} - \Pi_0(k,\omega)\, ,
 \ee

\noindent where $Z$ is the wavefunction renormalization constant,
namely the residue at the pole,

\be \label{Zres} Z^{-1} = 1+ \frac{1}{\pi} \mathcal{P}
\int_{0}^{\infty} 2 k_0
\frac{\mathrm{Im}\widetilde{\Sigma}^R_0(k_0,k)}{(k^2_0-\Omega^2_k)^2}~dk_0
\, ,  \ee

\noindent and $\Pi_0(k,\omega)$ is the real part of the twice
subtracted self-energy given by

\be \label{Pi0} \Pi_0(k,\omega) = -\frac{1}{\pi}
 (\omega^2-\Omega^2_k) ~\mathcal{P} \int_{0}^{\infty} 2 k_0
\frac{\mathrm{Im}\widetilde{\Sigma}^R_0(k_0,k)}{(k_0^2-\omega^2)(k^2_0-\Omega^2_k)^2}~dk_0
\ee


The two subtractions had been performed on the single particle
mass-shell. In a renormalizable theory the \emph{integral} in the
twice subtracted real part of the self energy $\Pi_0(k,\omega)$ is
ultraviolet \emph{finite} while the integral in $Z^{-1}$ is
logarithmically divergent. Furthermore the finite temperature parts
do not have primitive divergences since all the primitive
divergences are those of the zero temperature theory. We emphasize
that these expressions are still functions of the \emph{bare}
coupling and any potential divergences arising from coupling
renormalization have not yet been accounted for. The divergences
associated with coupling constant renormalization will be addressed
below.


Combining equations  (\ref{PoleT02}), (\ref{sub00}) and
(\ref{Zsub}), the spectral density (\ref{rho}) can be written in the
following form

\be\label{rhosub} \rho(k,\omega;T) = \frac{1}{\pi}\,
\frac{~\left[\mathrm{Im}\widetilde{\Sigma}^R(k,\omega;T)+2\omega\epsilon\right]}{\left[Z^{-1}(\omega^2-\Omega^2_k)-\widetilde{\Pi}(k,\omega;T)
\right]^2+\left[\mathrm{Im}\widetilde{\Sigma}^R(k,\omega;T)+2\omega\epsilon\right]^2}\;,
\ee

\noindent where

\be\label{PiofT} \widetilde{\Pi}(k,\omega;T)=
(\omega^2-\Omega^2_k)\Pi_0(k,\omega)+
\mathrm{Re}\widetilde{\Sigma}^R_T(\omega,k) \ee

Introducing the \emph{renormalized} real and imaginary part of the
self-energy as

\bea  && \widetilde{\Pi}_r(k,\omega;T) =   Z
\,\widetilde{\Pi}(k,\omega;T)\label{piren}\\
&& \mathrm{Im}\widetilde{\Sigma}^R_r(k,\omega;T) = Z\,
\mathrm{Im}\widetilde{\Sigma}^R(k,\omega;T)\label{imisigren} \eea

\noindent the spectral density (\ref{rhosub}) can be written  as

\be \rho(k,\omega;T) = Z\, \rho_r(k,\omega;T)\,, \label{rhoren}\ee

\noindent where

\be \label{rhoreno} \rho_r(k,\omega;T) = \frac{1}{\pi}\,
\frac{~\left[\mathrm{Im}\widetilde{\Sigma}^R_r(k,\omega;T)+2\omega\epsilon\right]}{\left[(\omega^2-\Omega^2_k)-\widetilde{\Pi}_r(k,\omega;T)
\right]^2+\left[\mathrm{Im}\widetilde{\Sigma}^R_r(k,\omega;T)+2\omega\epsilon\right]^2}\;.\ee

We note that at zero temperature the spectral density
$\rho_r(k,\omega;T=0)$ has unit residue at the single physical
particle pole.

Since both $\widetilde{\Pi}(k,\omega;T)$ and
$\mathrm{Im}\widetilde{\Sigma}^R(k,\omega;T)$ are proportional to
$g^2$, the renormalization of the real and imaginary part of the
self-energy in eqns. (\ref{piren}),(\ref{imisigren}) is tantamount
to the renormalization of the coupling constant\footnote{The
coupling \textit{g} in the Lagrangian already has the proper
renormalization of the (composite) operator $\mathcal{O}[\chi]$. }

\be \label{coupren} g_r = \sqrt{Z} g \ee

In terms of $g_r$, both $\widetilde{\Pi}_r(k,\omega;T)$ and
$\mathrm{Im}\widetilde{\Sigma}^R_r(k,\omega;T)$ are \emph{finite}
since the only counterterms necessary are those of the zero
temperature theory. Therefore the equilibrium distribution function
can be written solely in terms of renormalized quantities as follows

\be \label{Nfinalren} \mathcal{N}(k,T) =\int_{0}^{\infty}
\Bigg(\frac{\omega^2+\Omega^2_k}{2\,\Omega_k}\Bigg)\Bigg\{\left[1+2n(\omega)\right]\rho_r(k,\omega,T)-\rho_{r,c}(k,\omega,T=0)\Bigg\}\,{d\omega}\,
-\frac{1}{2} \;.\ee

This definition of the asymptotic distribution function is one of
the main results of this article.

\section{The model}\label{sec:model}

The results obtained in the previous section are general and as
mentioned above  the quantum kinetic effects that modify the
standard Boltzmann suppression of particle abundance in the medium
depend on the particular theory under consideration. To highlight
the main concepts in a specific scenario, we now consider a theory
of three interacting  real scalar fields with the following
Lagrangian density.

\be \label{Lag} \mathcal{L}= \frac{1}{2}
\partial_{\mu}\Phi\partial^{\mu}\Phi - \frac{1}{2} m^2 \Phi^2 +
\sum_{i=1}^2 \Bigg[\frac{1}{2}
\partial_{\mu}\chi_i\partial^{\mu}\chi_i - \frac{1}{2} M^2_i \chi^2_i
\Bigg]-g\Phi \,\chi_1\, \chi_2 +\mathcal{L}_{int}[\chi_1\,\chi_2]\ee

We will assume that the mutual  interaction between the fields
$\chi_1\,,\,\chi_2$  ensures that the fields $\chi_{1,2}$ are in
thermal equilibrium at a temperature $T=1/\beta$. A similar model
has been previously studied in ref.\cite{weldon} for an analysis
of the different processes in the medium.

The particles associated with the field $\Phi$ will be stable at
$T=0$ provided $m_P<M_1+M_2$, where $m_P$ is the zero temperature
pole mass of the $\Phi$ particles. In order to study the emergence
of a width for the particles of the field $\Phi$ to lowest order in
perturbation theory we will consider the case in which $M_1>m_P+M_2$
(or alternatively $M_2>m_P+M_1$) in this case the quanta of the
field $\chi_1$ can \emph{decay} into those of the field $\Phi$ and
$\chi_2$. Since the particles $1,2$ are in a thermal bath in
equilibrium the presence of the heavier species (here taken to be
that of the field $\chi_1$) in the medium results in a \emph{width}
for the excitations of field $\Phi$ through the process of decay of
the heavier particle  into the lighter scalars  and its
recombination, namely $\chi_1 \leftrightarrow \Phi + \chi_2$. As
will be seen in detail below the kinematics for this process is
similar to that for Landau damping in the case of massive particles
\cite{brapis}.

The relevant quantity is the self-energy of the field $\Phi$ which
we now obtain to one loop order $\mathcal{O}(g^2)$ in the Matsubara
representation. The one-loop self-energy is given by

\be\label{sigmafi}\Sigma(\nu_{n}, \vec{k})= -g^{2}\int
\frac{d^{3}\vec{p}}{(2\pi)^{3}}\frac{1}{\beta}\sum_{\omega_{m}}G_{\chi_{1}}^{(0)}(\omega_{m},
\vec{p})\,G_{\chi_{2}}^{(0)}(\omega_{m}+\nu_{n},
\vec{p}+\vec{k})\, ,
\end{equation}
where $ \omega_m,\nu_n$ are Bosonic Matsubara frequencies. It is
convenient to write the  Matsubara propagators  in the following
dispersive form

\begin{eqnarray}
G_{\chi_{1}}^{(0)}(\omega_{m}, \vec{p})&=& \int
dp_{0}\frac{\rho_{1}(p_{0}, \vec{p})}{p_{0}-i \omega_{m}}\, ,
\\
G_{\chi_{2}}^{(0)}(\omega_{m}+\nu_{n}, \vec{p}+\vec{k})&=& \int
dq_{0}\frac{\rho_{2}(q_{0}, \vec{p}+\vec{k})}{q_{0}-i
\omega_{m}-i\nu_{n}}\, ,
\\
\rho_{1}(p_{0},\vec{p})&=& \frac{1}{2\omega_{\vec{p}}^{(1)}}
[\delta(p_{0}-\omega_{\vec{p}}^{(1)})-\delta
(p_{0}+\omega_{\vec{p}}^{(1)})]\, ,
\\
\rho_{2}(q_{0},
\vec{p}+\vec{k})&=&\frac{1}{2\omega_{\vec{p}+\vec{k}}^{(2)}}[\delta
(q_{0}-\omega_{\vec{p}+\vec{k}}^{(2)})-\delta (q_{0}+\omega_{\vec{p}+\vec{k}}^{(2)})]\,, \\
\omega_{\vec{p}}^{(1)}&=& \sqrt{\vec{p}^{2}+M_{1}^{2}}\,; \,\,
\,\,\, \omega_{\vec{p}+\vec{k}}^{(2)} =
\sqrt{(\vec{p}+\vec{k})^{2}+M_{2}^{2}}\,.
\end{eqnarray}

This representation allows to carry out the sum over Matsubara
frequencies $\omega_{m}$ in a rather straightforward
manner\cite{kapusta,lebellac} which automatically leads to the
following dispersive representation of the self-energy
\begin{equation}\label{sigdis}
\Sigma(k,\nu_n) = -\frac{1}{\pi} \int_{-\infty}^{\infty} d\omega
\frac{\mathrm{Im} \widetilde{\Sigma}^R(k,\omega)}{\omega-i \nu_n}
\end{equation}

with the imaginary part of the self-energy given by

\begin{equation}\label{imsigrep}
\text{Im} \widetilde{\Sigma}^R(k,\omega) = \pi g^{2}\int
\frac{d^{3}\vec{p}}{(2\pi)^{3}}\int dp_{0} \int dq_{0}
\,[n(p_{0})-n(q_{0})]\,\rho_{1}(p_{0}, \vec{p})\,\rho_{2}(q_{0},
\vec{p}+\vec{k})\,\delta(\omega-q_{0}+p_{0})
\end{equation}

\noindent where $n(q)$ are the Bose-Einstein distribution
functions.  From the representation (\ref{specsigret}) the
retarded self-energy follows by analytic continuation, namely

\be \widetilde{\Sigma}^R(k,k_0) = \Sigma(k,\nu_n=k_0-i\epsilon)
\ee

The imaginary part of the  self energy can be written as a sum of
several different contributions, namely

\begin{equation}\label{imsigsplit}
\text{Im}\widetilde{\Sigma}^R_r(k,\omega;T)=
\sigma_{0}(k,\omega)+\sigma_{a}(k,\omega;T)+\sigma_{b}(k,\omega;T)\;,
\end{equation}

\noindent where $\sigma_{0}(k,\omega)$ is the zero temperature
contribution given by

\be\label{sig01} \sigma_{0}(k,\omega)  =  \frac{g^{2}}{32
\pi^{2}}\, \int
\frac{d^{3}\vec{p}}{\omega_{\vec{p}}^{(1)}\omega_{\vec{p}+\vec{k}}^{(2)}}\,
\Bigg[\delta\left(\omega-\omega_{\vec{p}}^{(1)}-\omega_{\vec{p}+\vec{k}}^{(2)}\right)-
\delta\left(\omega+\omega_{\vec{p}}^{(1)}+\omega_{\vec{p}+\vec{k}}^{(2)}\right)\Bigg],\ee

\noindent and $\sigma_{a}(k,\omega),\sigma_{b}(k,\omega)$ are the
finite temperature contributions given by

\be \label{siga1} \sigma_{a}(k,\omega;T)=  \frac{g^{2}}{32
\pi^{2}}\, \int
\frac{d^{3}\vec{p}}{\omega_{\vec{p}}^{(1)}\omega_{\vec{p}+\vec{k}}^{(2)}}\,
\Big[n(\omega_{\vec{p}}^{(1)})+n(\omega_{\vec{p}+\vec{k}}^{(2)})
\Big]\Bigg[
\delta\left(\omega-\omega_{\vec{p}}^{(1)}-\omega_{\vec{p}+\vec{k}}^{(2)}\right)-
\delta\left(\omega+\omega_{\vec{p}}^{(1)}+\omega_{\vec{p}+\vec{k}}^{(2)}\right)\Bigg]\,,\ee

\be \label{sigb1} \sigma_{b}(k,\omega;T)=  \frac{g^{2}}{32
\pi^{2}}\, \int
\frac{d^{3}\vec{p}}{\omega_{\vec{p}}^{(1)}\omega_{\vec{p}+\vec{k}}^{(2)}}\,
\Big[n(\omega_{\vec{p}+\vec{k}}^{(2)})-n(\omega_{\vec{p}}^{(1)})
\Big] \Bigg[
\delta\left(\omega-\omega_{\vec{p}}^{(1)}+\omega_{\vec{p}+\vec{k}}^{(2)}\right)-
\delta\left(\omega+\omega_{\vec{p}}^{(1)}-\omega_{\vec{p}+\vec{k}}^{(2)}\right)\Bigg]\,,\ee

\begin{figure}
\begin{center}
\includegraphics[height=3.5in,width=3.5in,keepaspectratio=true]{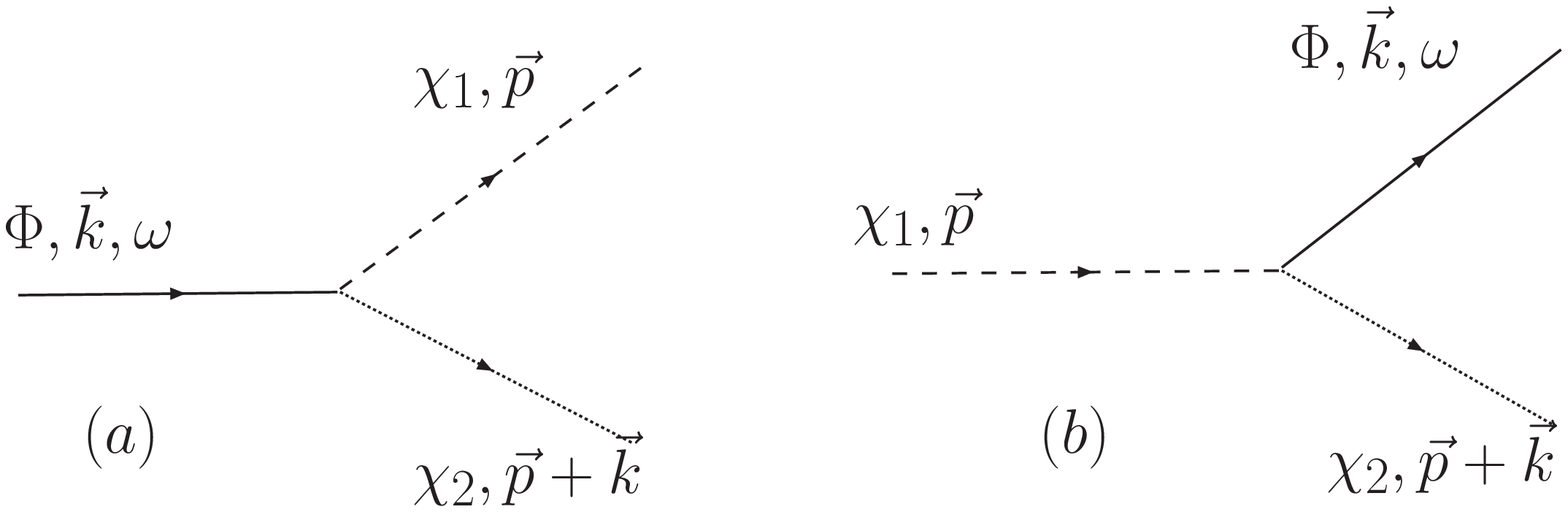}
\caption{Processes contributing to
$\sigma_0(k,\omega),\sigma_a(k,\omega)$ (a) and to
$\sigma_b(k,\omega)$ (b). The inverse processes are not
shown.}\label{fig:decays}
\end{center}
\end{figure}

The processes that contribute to $\sigma_0(k,\omega)$ and
$\sigma_a(k,\omega)$ are $\Phi\leftrightarrow \chi_1 \,\chi_2$
while the processes that contribute to $\sigma_b(k,\omega)$ are
$\chi_{1,2} \leftrightarrow \Phi \, \chi_{2,1}$ depicted
schematically in fig. (\ref{fig:decays})

The details of the calculation of the different contributions are
relegated to the appendix. The result is summarized as  follows:

\be\label{sig0} \sigma_{0}(k,\omega)  =  \frac{g^{2}}{16 \pi Q^2 }\,
\text{sign}(\omega)\, \Theta[Q^2-(M_{1}+M_{2})^{2}\,]\,
\Big[(Q^2)^2-2Q^2(M^2_1+M^2_2)+(M^2_1-M^2_2)^2
\Big]^\frac{1}{2}~~;~~Q^2=\omega^{2}-k^{2} \ee

We have explicitly displayed the fact that the zero temperature
contribution to the imaginary part is manifestly Lorentz invariant
and solely a function of the invariant mass $Q^2=\omega^2-k^2$. The
finite temperature contributions are

\be\label{siga}\sigma_{a}(k,\omega;T)  =  \frac{g^{2}}{16 \pi
k\,\beta }\, \text{sign}(\omega)\,
\Theta[Q^2-(M_{1}+M_{2})^{2}\,]\,\Bigg[ \ln \left( \frac{
1-e^{-\beta \omega_{p}^{+}} }{ 1-e^{-\beta
\omega_{p}^{-}}}\right)+ M_1 \leftrightarrow M_2 \Bigg]\ee

\be\label{sigb}\sigma_{b}(k,\omega;T)  =  \frac{g^{2}}{16 \pi
k\,\beta }\, \text{sign}(\omega)\,
\Theta[(M_{1}-M_{2})^{2}-Q^2\,]\,\Bigg[ \ln \left( \frac{
1-e^{-\beta |\omega_{p}^{-}|} }{ 1-e^{-\beta
|\omega_{p}^{+}|}}\right)+ M_1 \leftrightarrow M_2 \Bigg]\ee

\noindent where

\be\label{opm} \omega_p^{\pm} =
\frac{|\omega|}{2Q^2}(Q^2+M^2_1-M^2_2)\pm
\frac{k}{2Q^2}\Bigg[(Q^2+M^2_1-M^2_2)^{2}-4Q^2M_{1}^{2}\Bigg]^\frac{1}{2}~~;~~Q^2=\omega^{2}-k^2.
\ee

\begin{figure}[ht!]
\begin{center}
\includegraphics[height=3in,width=3in,keepaspectratio=true]{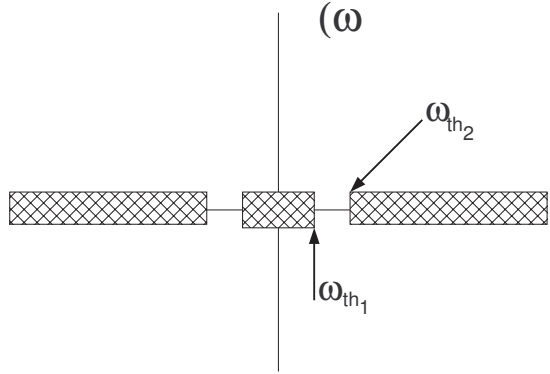}
\caption{Spectral density $\rho(k,\omega,T)$ for
$m^2_P<(M_{1}-M_{2})^{2}$. The shaded areas are the multiparticle
cuts with thresholds $\omega_{th 1}=\sqrt{k^2+(M_1-M_2)^2}$ and
$\omega_{th 2}=\sqrt{k^2+(M_1+M_2)^2}$. The single particle poles at
$\Omega^2_k=k^2+m^2_P$ moved off the real axis into an unphysical
sheet. \label{fig:rhoT} }
\end{center}
\end{figure}

The real part of the self energy is obtained from the dispersive
form (\ref{specsigret}) and can be separated into a zero temperature
and a finite temperature part as follows

\be \label{SEsplit} \mathrm{Re}\widetilde{\Sigma}^R(k,\omega;T ) =
 \mathrm{Re}\widetilde{\Sigma}^R_{0}(k,\omega ) +
\mathrm{Re}\widetilde{\Sigma}^R_{T}(k,\omega;T )  \ee

\noindent with

\bea \label{SES} \mathrm{Re}\widetilde{\Sigma}^R_{0}(k,\omega ) &
= & -\frac{1}{\pi} \mathcal{P} \int_{-\infty}^{\infty}
\frac{\sigma_0(k_0,k)}{k_0-\omega}~dk_0
\label{SETE0}\\
\mathrm{Re}\widetilde{\Sigma}^R_{T}(k,\omega;T )  & = &
-\frac{1}{\pi} \mathcal{P} \int_{-\infty}^{\infty}
\frac{\sigma_a(k_0,k;T)+\sigma_b(k_0,k;T)}{k_0-\omega}~dk_0\label{SETT}
\eea

\noindent where $\mathcal{P}$ stands for the principal part.  We
note that both $\sigma_0(k,\omega)$ and $\sigma_a(k,\omega)$ feature
the standard two particle threshold above the light cone at the
invariant mass $Q^2=(M_{1}+M_{2})^2$ whereas the finite temperature
contribution $\sigma_b(k,\omega)$ has support for invariant mass
$Q^2 \leq (M_{1}-M_{2})^{2}$ even \emph{below the light cone} and
vanishes at $T=0$. In the case of massless particles in the loop
this contribution is below the light cone and is identified with
Landau damping\cite{kapusta,brapis,lebellac}. In particular at zero
temperature the isolated poles are at $Q^2=m^2_P$, hence if
$m^2_P<(M_{1}-M_{2})^{2}$ the physical particle pole is embedded in
the multiparticle continuum and moves off the real axis onto the
second (or higher) Riemann sheet in the complex frequency plane.
Because of this the physical particle acquires a width. The spectral
density for the case $m^2_P<(M_{1}-M_{2})^{2}$ is depicted in fig.
(\ref{fig:rhoT})

\subsubsection{Zero temperature: $\Omega_k;Z;\mathcal{C}_k$:}

 Using that $\sigma_0(k_0,k)$ is odd in $k_0$ and that it is
solely a function of the invariant $P^2= k^2_0-k^2$ for $k_0 >0$, it
is straightforward to find the following manifestly  Lorentz
invariant result

\be \label{SigRe0} \mathrm{Re}\widetilde{\Sigma}^R_{0}(k,\omega )
= -\frac{1}{\pi} \mathcal{P} \int_{(M_1+M_2)^2}^{\infty}
\frac{\sigma_0(P^2)}{P^2-Q^2}~dP^2~~;~~Q^2=\omega^2-k^2 \ee

\noindent where we have explicitly exhibited the two particle
threshold in the lower limit. Lorentz invariance requires that the
single particle pole features the dispersion relation $\Omega^2_k =
k^2+ m^2_P$, and so the equation that determines the single particle
physical poles, namely eqn. (\ref{PoleT0}) is given by

\be\label{polito} m^2_P-m^2
-\mathrm{Re}\widetilde{\Sigma}^R_{0}(Q^2= m^2_P) =0 \ee

From the results of the previous section (see eqn. (\ref{Zres}))
the wave function renormalization constant is  given by

\be \label{Zresmod} Z^{-1} = 1+ \frac{1}{\pi} \mathcal{P}
\int_{(M_1+M_2)^2}^{\infty} ~ \frac{\sigma_0(P^2)}{(P^2-m^2_P)^2}~
dP^2 \, , \ee

Separating the residue at the physical particle pole and following
the steps described in section (\ref{renormalization}) the
renormalized spectral density (\ref{rhoren},\ref{rhoreno}) at zero
temperature can now be written in the following simple form

\be\label{rhomo} \rho_r(k,\omega;T=0) = \frac{1}{\pi}\,
\frac{\left[\sigma_{0,r}(Q^2)+2\omega\epsilon\right]}{\left[
(Q^2-m^2_P)\Big(1-\Pi_{0,r}(Q^2) \Big)
\right]^2+\left[\sigma_{0,r}(Q^2)+2\omega\epsilon\right]^2} =
\mathrm{sign}(\omega)\,
\delta(\omega^2-\Omega^2_k)+\rho_{c,r}(k,\omega;T=0) \ee

\noindent where $\sigma_{0,r}(k,\omega)=Z \sigma_0(k,\omega)$ is
given by the expression (\ref{sig0}) but with the coupling
constant replaced by the renormalized coupling $Zg^2=g^2_r$

The continuum contribution is given by

\be\label{rhomocont} \rho_{c,r}(k,\omega;T=0) = \frac{1}{\pi}\,
\frac{\sigma_{0,r}(Q^2)}{\left[ (Q^2-m^2_P)\Big(1-\Pi_{0,r}(Q^2)
\Big) \right]^2+\Big[\sigma_{0,r}(Q^2)\Big]^2}\;, \ee

\noindent with

\be \label{Pi0ren} \Pi_{0,r}(Q^2) = -\frac{1}{\pi}
 (Q^2-m^2_P) ~\mathcal{P} \int_{(M_1+M_2)^2}^{\infty}
\frac{\sigma_{0,r}(P^2)}{(P^2-Q^2)(P^2-m^2_P)^2}~dP^2 \ee

\noindent where we have made explicit the two particle threshold
in the lower limit of the integral.

The \emph{exact} expression for $Z$ given by the sum rule
(\ref{ZT02}) coincides with $Z$ given by  eqn. (\ref{Zresmod}) to
lowest order in perturbation theory ($\mathcal{O}(g^2)$).

Up to $\mathcal{O}(g^2)$  we can  neglect $\sigma_0(k,\omega)$ as
well as $\Pi_{0,r}(k,\omega)$ in the denominator of the continuum
contribution (\ref{rhomocont}) because $Q^2 \geq (M^2_1+M^2_2)>
m^2_P$ and the denominator is never perturbatively small. Therefore
to leading order in the coupling we can approximate

\be\label{rhomoap} \rho_{c,r}(k,\omega;T=0) \simeq \frac{1}{\pi}\,
\frac{\sigma_{0,r}(Q^2)}{(Q^2-m^2_P)^2}\;. \ee

The renormalized spectral function at finite temperature can be
separated into the contributions from the different multiparticle
cuts,

\be \rho_r(k,\omega,T) =
\rho_{I,r}(k,\omega,T)+\rho_{II,r}(k,\omega,T) \ee

\noindent where the contribution with support above the two
particle cut is

\be\label{rhoI} \rho_{I,r}(k,\omega;T) = \frac{1}{\pi}\,
\frac{\left[\sigma_{0,r}(k,\omega)+\sigma_{a,r}(k,\omega;T)\right]}{\left[
(Q^2-m^2_P)\Big(1-\Pi_{0,r}(Q^2)
\Big)-\mathrm{Re}\widetilde{\Sigma}^R_{T,r}(k,\omega;T )
\right]^2+\Big[\sigma_{0,r}(Q^2)+\sigma_{a,r}(k,\omega;T)\Big]^2}\ee

\noindent  and that which has support below the light cone given
by

\be\label{rhoII} \rho_{II,r}(k,\omega;T) = \frac{1}{\pi}\,
\frac{\sigma_b(k,\omega;T)}{\left[
(Q^2-m^2_P)\Big(1-\Pi_{0,r}(Q^2)
\Big)-\mathrm{Re}\widetilde{\Sigma}^R_{T,r}(k,\omega;T )
\right]^2+\Big[\sigma_{b,r}(k,\omega;T)\Big]^2}\ee

\noindent where again the renormalized quantities are obtained
from the unrenormalized ones by replacing $g\rightarrow g_r=Zg$.

Since $\rho_{I,r}(k,\omega)$ has support only for
$|\omega|>\sqrt{k^2+(M^2_1+M^2_2)}$ its denominator is never
perturbatively small, therefore to leading order
$\mathcal{O}(g^2)$
 in the perturbative expansion it can be approximated by

\be\label{rhoIapp} \rho_{I,r}(k,\omega;T) \simeq \frac{1}{\pi}\,
\frac{\left[\sigma_{0,r}(k,\omega)+\sigma_{a,r}(k,\omega;T)\right]}{
(\omega^2-k^2-m^2_P)^2}\ee

For $\rho_{II,r}(k,\omega)$ we must keep the full expression because
for $m^2_P<(M_1-M_2)^2$ the denominator becomes perturbatively small
for $\omega^2 \sim k^2+m^2_P$. Therefore the final expression for
the asymptotic distribution function (\ref{Nfinalren}) to leading
order in the coupling ($\mathcal{O}(g^2)$) is given by

\be\label{Nasifin}\mathcal{N}(k,T) =
\mathcal{N}_I(k;T)+\mathcal{N}_{II}(k,T)\ee

\noindent where the different contributions reflect the different
multiparticle cuts, namely

\bea  \mathcal{N}_{II}(k,T) & = & \int_{0}^{\infty}
\Bigg(\frac{\omega^2+\Omega^2_k}{2\,\Omega_k}\Bigg)\Bigg\{\left[1+2n(\omega)\right]\rho_{II,r}(k,\omega,T)\Bigg\}\,{d\omega}\,
-\frac{1}{2} \label{NII} \\
 \mathcal{N}_{I}(k,T) & = & \frac{2}{\pi}\int_{\omega_{th}(k)}^{\infty}
\Bigg[\frac{\omega^2+\Omega^2_k}{2\,\Omega_k
(\omega^2-\Omega^2_k)^2}\Bigg]\Bigg\{n(\omega)\Big[\sigma_{0,r}(k,\omega)+\sigma_{a,r}(k,\omega;T)
\Big]+\frac{1}{2}\sigma_{a,r}(k,\omega;T)
\Bigg\}\,{d\omega}\label{NI} \, ,\eea

\noindent where
$\omega_{th}(k)=\left[k^2+(M_1+M_2)^2\right]^{\frac{1}{2}}$ is the
two particle cut.

\subsection{The approach to equilibrium:}\label{sec:equilib}
Before we study the asymptotic distribution function we address the
approach to equilibrium.  The time evolution of the (interpolating)
number operator $N_k(t)$ given by eqns. (\ref{finumber}-\ref{funF})
is completely determined by the real time evolution of the solution
$f_k(t)$ given by eqn. (\ref{frho}). For $m_P < |M_1-M_2|$ the
particle acquires a width through the two body decay of the heavier
particle in the bath and the particle pole is now embedded in the
continuum for $Q^2<(M_1-M_2)^2$, which is the relevant part of the
spectral density is $\sigma_b(k,\omega,T)$ given in eqn.
(\ref{sigb}). In the Breit Wigner approximation, the spectral
density is given by eqns. (\ref{rhoBW},\ref{QPole},\ref{ZT}) with

\be\label{gamb} \Gamma_k(T)= \mathcal{Z}_k
\,\frac{\sigma_b(k,\mathcal{W}_k(T),T)}{2\,\mathcal{W}_k(T)}\ee

\begin{figure}
\begin{center}
\includegraphics[height=3in,width=3in,keepaspectratio=true]{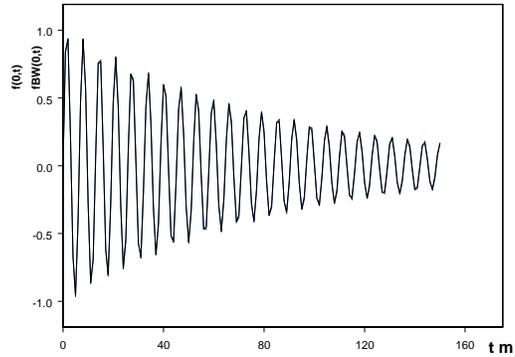}
\caption{The functions $f_{k=0}(t)$  and $f^{BW}_{k=0}(t)$ vs
$t\,m_P$ for $g^2/(16\pi^2 m^2_P)=0.01$
$M_1=4m_P\,;\,M_2=m_P\,;\,T=10m_P$. For these values of the
parameters we find numerically:
$\mathcal{Z}_0(T)=0.998\;,\;\mathcal{W}_0(T)= 0.973 m_P\;,\;
\Gamma_0(T)= 0.012 m_P$. The exact solution  and the  Breit-Wigner
approximation are indistinguishable. } \label{fig:f0oft}
\end{center}
\end{figure}

The real time evolution of the solution $f_k(t)$ in the Breit-Wigner
approximation is given by eqn. (\ref{RTTf}). Figure
(\ref{fig:f0oft}) displays both the \emph{exact} solution
(\ref{frho}) and the Breit-Wigner approximation (\ref{RTTf}) for
$k=0$. The exact and approximate solutions are indistinguishable
during the time scale of the numerical evolution as gleaned from
this figure.

The asymptotic long time evolution is determined by the behavior
of the spectral density near the thresholds and is typically of
the form of a power law\cite{threshold}. However, such asymptotic
behavior sets in at very long times, beyond the regime in which
our numerical study is trustworthy. It is numerically exceedingly
difficult to extract the exponential relaxation from the power
laws that dominate at asymptotically long time because the
amplitude becomes very small in the weak coupling case.

The main conclusion is that the distribution function approaches
thermalization and becomes insensitive to the initial conditions
for time scales $t > \tau_k = 1/2\Gamma_k(T)$, where $\Gamma_k(T)$
is the quasiparticle relaxation rate.

\subsection{The asymptotic distribution function:}

In the Breit-Wigner approximation and assuming a very narrow
resonance near the physical particle pole

\be \rho_{II,r}(k,\omega,T) \sim \frac{1}{\pi}\,
\frac{\mathrm{sign}(\omega)\,\Gamma_k}{(\omega^2-\Omega^2_k)^2+\Gamma^2_k}
\sim \mathrm{sign}(\omega) \delta(\omega^2-\Omega^2_k)
\label{narrowBW}\ee

\noindent where in the second term on the right hand side  the width
has been neglected by assuming a very narrow resonance at
$\Omega_k$. Therefore in this narrow width approximation one would
expect that the different contributions are given by

\be \mathcal{N}_{II}(k,T)\sim n(\Omega_k)~~;~~\mathcal{N}_I(k,T) =
\mathcal{O}(g^2) \ee

\noindent where $n(\Omega_k)$ is the Boltzmann distribution function
for the stable particle. This rather simple analysis would lead to
the conclusion that the corrections to the equilibrium abundance are
perturbatively small.

However, even for weakly coupled theories we expect this simple
argument to be incorrect both in the high and low temperature
regimes. The main reason for this expectation is that the
approximation (\ref{narrowBW})  suggests that this argument neglects
the fact that the spectral density has support for frequencies
\emph{smaller} than  the position of the physical particle pole
(namely for $|\omega| \neq \Omega_k$). From the expression
(\ref{NII}) it is clear that the region of small $\omega$ will lead
to a substantial correction since for $\omega \ll T$ the
Bose-Einstein distribution function in (\ref{NII}) becomes
$n(\omega) \sim T/\omega >> 1$, thus the region of $|\omega| <
\Omega_k$ and in particular $|\omega| \ll T$ gives a non-trivial
contribution to the abundance. The region of spectral density for
$|\omega| > \Omega_k$ will yield a much smaller, but non-negligible
contribution. Furthermore in the high temperature limit $T \gg
k,m_P,M_{1,2}$ the width is expected to become large. This can be
gleaned from the expression for $\sigma_b(\omega,k,T)$ in eqn.
(\ref{sigb1}), which for $T \gg \omega^{1,2}_{p}$ is proportional to
$T$. This is clearly a statement that at high temperatures there is
a large population of heavy particles which results in a larger
number of processes $\chi_1 \leftrightarrow \Phi \, \chi_2$ in the
medium, thereby increasing the width of the particle $\Phi$. As the
width of the spectral density near the physical particle pole
increases, the spectral density has larger support in the small
$\omega$ region, thereby increasing the off-shell contributions to
the abundance. These arguments will be confirmed both analytically
and numerically below.

We now study numerically and analytically the asymptotic
distribution function to assess precisely the magnitude and origin
of the corrections to the equilibrium abundance. The parameter space
is fairly large, thus we consider separately the cases of small
momenta $k \ll m_P,M_1,M_2,T$ and the case of large momenta $k \gg
m_P,M_1,M_2,T$ choosing the unit of energy to be the zero
temperature pole mass of the particle, $m_P$ and keeping the value
of the masses of the heavy fields fixed with $M_1>M_2+m_P$.

\subsubsection{$k =0$}

The limit $k=0$ of the spectral density can be easily obtained from
the expressions given above (\ref{sig0}-\ref{sigb}). Of particular
importance is the high temperature limit of $\sigma_b(0,\omega,T)$
since this contribution to the spectral density determines the width
of the spectral function near the physical particle pole
$\Gamma_0(T)$ given by eqn. (\ref{gamb}).

A straightforward calculation leads to the following result in the
limit $T \gg m_P\,,\,M_{1,2}$,

\be\label{widthK0} \frac{\sigma_b(0,m_P,T)}{2\,m_P} = \frac{g^2
T}{8\pi^2} \, \frac{\Big[m^4_P+(M^2_1-M^2_2)^2-2m^2_P(M^2_1+M^2_2)
\Big]^{\frac{1}{2}}}{\Big[ (M^2_1-M^2_2)^2-m^4_P\Big]}\ee

\begin{figure}
\begin{center}
\includegraphics[height=3in,width=3in,keepaspectratio=true]{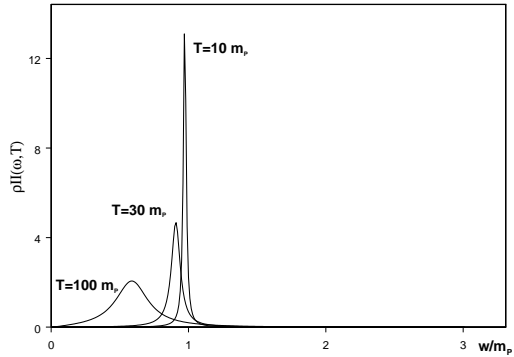}
\caption{The spectral density $\rho_{II}(k=0,\omega,T)$ vs
$\omega/m_P$ for $g^2/(16\pi^2 m^2_P)=0.01$ $M_1=4m_P\,;\,M_2=m_P$
and $T/m_p=10,30,100$ respectively.} \label{fig:rhok0}
\end{center}
\end{figure}

We note that this expression for the width is \emph{classical}
since restoring $g^2 \rightarrow g^2 \hbar\,;\, T \rightarrow
T/\hbar$ the expression above is independent of $\hbar$. This is a
consequence of the fact that the high temperature limit is
completely determined by the Rayleigh-Jeans part of the
Bose-Einstein distribution function. As a result when the
temperature is much larger than all mass scales, the width is
proportional to $T$ and the spectral density becomes wider,
enhancing the off-shell contributions. Fig. (\ref{fig:rhok0})
displays the spectral density for several values of the
temperature highlighting the broadening for large temperature. It
is clear from this figure that at very high temperatures
perturbation theory breaks down in this model since the width can
become comparable to the physical mass or the position of the
pole. This situation has been previously noticed in a scalar field
theory at high temperatures,  and a finite temperature
renormalization group was introduced to provide a non-perturbative
resummation\cite{hotRG}.

Restricting ourselves to the regime in temperature within which
perturbation theory is still reliable, namely for $\Gamma_0(T) \ll
m_P$,  we study the departure of the distribution function from the
Bose-Einstein form (for $k=0$) numerically.

\begin{figure}
\begin{center}
\includegraphics[height=3in,width=3in,keepaspectratio=true]{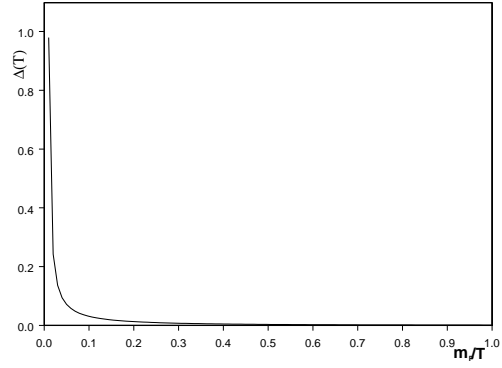}
\caption{The ratio $\Delta(T)=(\mathcal{N}(0,T)-n(m_P))/n(m_P)$ vs
$m_P/T$ for $g^2/(16\pi^2 m^2_P)=0.01$ $M_1=4m_P\,;\,M_2=m_P$.}
\label{fig:delta}
\end{center}
\end{figure}

Figure (\ref{fig:delta}) displays the quantity

\be\label{Delt} \Delta(T) =
\frac{\mathcal{N}(k=0,T)-n(m_P)}{n(m_P)}\ee

\noindent  for a weakly coupled case in the range of temperatures
$1 \leq T/m_p \leq 20$ for $M_1=4m_P\,,\,M_2=m_P$ within which we
find numerically that $\Gamma_0(T)/m_P \leq 0.1$ which we use as a
reasonable criterion for the validity of perturbation theory (see
fig. \ref{fig:rhok0}).

This figure clearly indicates that even within the high temperature
regime where perturbation theory is reliable and the spectral
density still features a rather narrow Breit-Wigner peak, there are
substantial departures from  the Bose-Einstein form in the
equilibrium distribution function. At low temperatures fig.
(\ref{fig:delta}) clearly displays an exponential suppression and
the distribution function essentially becomes the Bose-Einstein
distribution. In this limit the width is extremely small and the
spectral density is almost a delta function on the physical particle
mass shell, and the off-shell effects are perturbatively small.

\subsubsection{$k\gg T,m_P,M_{1,2}$}

\begin{figure}[h!]
\begin{center}
\includegraphics[height=3in,width=3in,keepaspectratio=true]{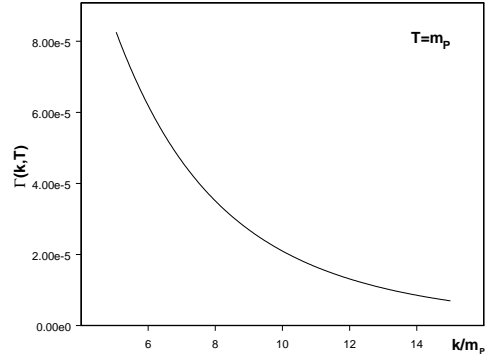}
\caption{The width $\Gamma(k,T)$ in units of $m_P$  vs $k/m_P$ for
$k \gg T$   for $T=m_P$ for $g^2/(16\pi^2 m^2_P)=0.01$
$M_1=4m_P\,;\,M_2=m_P$.} \label{fig:gamma}
\end{center}
\end{figure}

In the limit of large momenta several interesting features emerge:
i) the width of the spectral density becomes very small, this is a
consequence of the fact that there are very few heavy states for
large momenta in the heat bath if the momentum is large. The width
as a function of $k$ is depicted in fig. (\ref{fig:gamma}), which
displays clearly this behavior. ii) As a function of the variable
$\omega$, the position of the peak in the spectral density becomes
closer to the threshold for $k\gg m_P,M_{1,2}$. As a result of
both these effects the spectral distribution becomes strongly
peaked near threshold and the threshold moves to larger values of
the frequency, thus leaving behind a larger region of the spectra
off-shell for frequencies \emph{smaller} than the position of the
peak. The spectral density while small away from the peak is,
however, non-vanishing and the fact that there is now a larger
region in frequency $\omega$ below the (narrow) peak, brings about
a competition of scales as can be understood from the following
argument. The very narrow peak (almost a delta function at
$\omega=\Omega_k \sim k$) leads to a contribution
$\mathcal{N}_{II}(k,T)\sim n(\Omega_k)$, which for $k\gg T$ is
$\ll 1$. This on-shell contribution  competes against the
off-shell contributions from integrating the spectral density for
$\omega < \Omega_k$ which is also very small because
$\sigma_b(k,\omega,T)/\Omega^2_k \ll 1$ but for $\omega \ll T$ is
multiplied by the Bose enhancement factor $\sim T/\omega$. The
competition between the ``on-shell'' contribution $n(\Omega_k)$
and the off-shell contributions is studied numerically.

\begin{figure}[h!]
\begin{center}
\includegraphics[height=3in,width=3in,keepaspectratio=true]{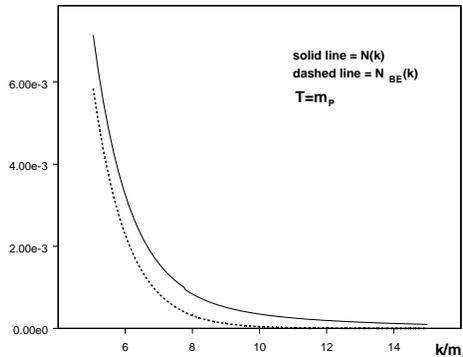}
\caption{The distribution functions $\mathcal{N}(k,T)$ given by eqn.
(\ref{Nasifin}) (solid line) vs. the Bose-Einstein distribution
$N_{BE}(k,T)$ (dashed line) as a function of $k/m_P$ for $T=m_P$;
$g^2/(16\pi^2 m^2_P)=0.01$; $M_1=4m_P\,;\,M_2=m_P$.}
\label{fig:occupa}
\end{center}
\end{figure}

Fig. (\ref{fig:occupa}) displays both the Bose-Einstein
distribution function $n(\Omega_k)$ and the asymptotic
distribution function $\mathcal{N}(k,T)$ in the limit $k \gg T,
m_P, M_{1,2}$.

It is clear from this figure that while the distribution function
$\mathcal{N}(k,T)$ is strongly suppressed for $k\gg T$ it is
\emph{larger} than the Bose-Einstein distribution. The main reason
for this enhancement is precisely the competition mentioned above,
namely the position of the peak in the spectral density moves
towards threshold which for large $k$ corresponds to large values
of the frequency $\omega$. Therefore there is a large region in
which the  spectral density is very small but non-vanishing for
$\omega < \Omega_k$. Clearly the part of the spectral density with
support for $\omega > \Omega_k$ yields a much smaller contribution
to the distribution function. Furthermore, for  $\omega \ll T $
the factor $n(\omega) \sim T/\omega \gg 1$ which enhances further
the off-shell contributions.

These results in the different regimes can be summarized as
follows:

\begin{itemize}
\item{In the high temperature regime the larger abundance of heavier particles in the bath leads to a
broadening of the spectral density. This broadening in turn results
in a larger off-shell contribution to the abundance
$\mathcal{N}(k,T)$ and an enhancement of the distribution function
over the Bose-Einstein result. The off-shell region of small
frequency yields a substantial contribution because of the factor
$n(\omega) \sim T/\omega $ in (\ref{Nfinalren}). In the model
considered perturbation theory breaks down at high temperature and
the imaginary part on-shell becomes classical. This situation is
akin to the case of a self-interacting bosonic field theory studied
in ref. \cite{hotRG}. A  high temperature renormalization group
resummation program such as in ref.\cite{hotRG} may be required to
provide a non-perturbative resummation.  }

\item{ For momenta much larger than the mass scales and the
temperature there is also a large enhancement of the distribution
function $\mathcal{N}(k,T)$ over the Bose-Einstein result. In this
case the spectral density features a very narrow resonance near
the position of the physical pole at $\omega \simeq \Omega_k$,
which however moves closer to threshold. For large $k$ the
off-shell region of support of the spectral density becomes larger
and though the spectral density is strongly suppressed, the
off-shell contribution from the region $\omega < \Omega_k$
competes with the contribution from the on-shell pole, namely the
Bose-Einstein distribution function $n(\Omega_k)$ because for $k>>
T$  $n(\Omega_k) \ll 1$. The off-shell contribution from the
region $\omega << \Omega_k$ is comparable to or larger than
$n(\Omega_k)$ for $k \gg T$ and is enhanced in the region $\omega
\ll T$ by the factor $n(\omega)\sim T/\omega$. }

\end{itemize}

While these results may be particular to the model studied, we
expect most of these  features to be robust and fairly general. In
particular at high temperature it is physically reasonable to expect
a thermal broadening of the spectral density either from collisions,
 many-body decays or Landau damping as in the case studied here. Broadening of the
spectral function yields a larger contribution from the small
$\omega$ region which is enhanced further by the factor $n(\omega)
\sim T/\omega$ for $\omega \ll T$. Therefore a substantial departure
of the distribution function $\mathcal{N}(k,T)$ from the
Bose-Einstein distribution is expected at high temperature. A
possible breakdown of perturbation theory in the high temperature
regime may require the implementation of a non-perturbative
resummation procedure akin to that introduced in ref.\cite{hotRG}.
At low temperatures, much lower than the mass and momentum scales a
departure from simple Bose-Einstein is also expected. In this case
even though the spectral function features a sharp and narrow peak
at a position very near the physical particle pole, the
Bose-Einstein distribution function is very small. Hence the
off-shell region $\omega << \Omega_k$ of the spectral function will
lead to a substantial contribution which is further enhanced by the
factor $n(\omega) \sim T/\omega$ for $\omega \ll T$. Again because
the temperature is much smaller than any of the scales, the spectral
density will be exponentially suppressed off shell and the
equilibrium abundance will reflect this suppression, but just as in
the case studied here, may still be larger than the simple
Bose-Einstein abundance.  Of course our study within this particular
model serves only as a guidance and the details of the enhancement
will depend on the theory under consideration, but the main lesson
learned here is that the off-shell, small frequency region of the
spectral density yields a substantial contribution to the
equilibrium abundance in interacting theories.

\section{Boltzmann kinetics in renormalized perturbation
theory}\label{sec:boltzkin}

It is important to understand the origin of the differences between
the quantum kinetic equation for the distribution function
(\ref{finumber}) and the usual quantum Boltzmann equation. Therefore
in this section we provide a derivation of the quantum Boltzmann
equation in \emph{renormalized perturbation theory} to highlight the
origin of the different equilibrium abundances. We assume that the
bath is in equilibrium just as we did in our derivation of the
effective action and  the time evolution for the distribution
function in the previous sections.

The quantum Boltzmann equation is a differential equation for the
single particle distribution function. However, as we have discussed
in detail above, the \emph{physical} particles have mass $m_P$ and
the Heisenberg field operators create physical particles out of the
vacuum with an amplitude determined by the wave function
renormalization. Therefore in order to account for the mass and wave
function renormalization, and to obtain the kinetic Boltzmann
equation for the physical particles it is convenient to re-write the
Lagrangian by introducing counterterms, namely

\bea \label{Lagcount} \mathcal{L}= && \frac{1}{2}
\partial_{\mu}\Phi_r\partial^{\mu}\Phi_r - \frac{1}{2} m^2_P \Phi^2_r +
\sum_{i=1}^2 \Bigg[\frac{1}{2}
\partial_{\mu}\chi_i\partial^{\mu}\chi_i - \frac{1}{2} M^2_i \chi^2_i
\Bigg]-g_r\Phi_r \,\chi_1\, \chi_2
+\mathcal{L}_{count}+\mathcal{L}_{int}[\chi_1\,;\,\chi_2]\label{lagrac}\\
\mathcal{L}_{count}=&&\frac{1}{2}(Z-1)
\partial_{\mu}\Phi_r\partial^{\mu}\Phi_r - \frac{1}{2} \Delta m^2
\Phi^2_r \label{Lcount}\eea

\noindent where $g_r=\sqrt{Z}g$ and we assume that the
renormalization aspects of the fields $\chi_{1,2}$ had already been
included in $\mathcal{L}_{int}[\chi_1\,;\,\chi_2]$ .  The
counterterms in (\ref{Lcount}) are treated systematically in
perturbation theory along with the cubic interaction. Note that
$Z-1\,,\,\Delta m^2$ are both of $\mathcal{O}(g^2)$.

The renormalized field $\Phi_r$ is expanded in terms of creation and
annihilation operators of physical Fock states,

\bea  \Phi_r(\vec{x},t) & =  & \frac{1}{\sqrt{V}} \sum_{\vk}
\Phi_{\vk}(t) e^{i\vk\cdot\vec{x}} \label{expafi} \\
\Phi_{\vk}(t) & = & \frac{1}{\sqrt{2\Omega_k}}
\left[a_{\vk}\,e^{-i\Omega_k\,t}+a^{\dagger}_{-\vk}\,e^{i\Omega_k\,t}\right]\label{ani}\eea

\noindent and a similar expansion for the bath fields
$\chi_1,\chi_2$ in terms of creation and annihilation operators and
the corresponding frequencies $\omega_k^{(1,2)}$.  The total
interaction Lagrangian is

\be L_{int} = \frac{g_r}{\sqrt{V}}\sum_{\vk}\sum_{\vec{p}} \Phi_k\,
\chi_{1,\vec{p}} \,\chi_{2,-\vec{p}-\vk}+L_{count} \ee

\noindent where $L_{count}$ is the counterterm Lagrangian. The
kinetic Boltzmann equation for the occupation number of the Fock
quanta of the field $\Phi$ is

\be \frac{dN_k}{dt} =
\frac{dN_k}{dt}\Bigg|_{gain}-\frac{dN_k}{dt}\Bigg|_{loss}\ee

The gain and loss terms are obtained by calculating the transition
\emph{probabilities} per unit time for processes that lead to the
increase (gain) and decrease (loss) the occupation number, namely
$dN_k(t)/dt=dP_k/dt$. Within the framework of the kinetic
description such calculation is carried out by implementing
Fermi's Golden rule. The processes that lead to the increase or
decrease of the population are read-off the interaction and energy
conservation emerges  as a consequence of taking the long time
limit as is manifest in Fermi's Golden rule. The cubic interaction
term in $L_{int}$ gives rise to several different processes which
are gleaned by expanding the product in terms of the creation and
annihilation operators of all the fields involved. The different
phases that enter in such terms determine the energy conservation
delta functions in Fermi's Golden rule. Some of the processes are
depicted in fig. (\ref{fig:decays}). When $m_P<M_1,M_2$ the quanta
of the field $\Phi$ cannot decay into those of the bath fields,
however if $M_1>M_2+m_P$ (or $M_2>M_1+m_P$) the heavier bath field
can decay into particles of $\Phi$ therefore increasing the
population. This process is depicted in fig.
(\ref{fig:decays}-(b)). The inverse process contributes to the
loss term.  Let us consider the case $M_1>M_2+m_P$ (the case
$M_2>M_1+m_P$ is similar by $M_2\leftrightarrow M_1$). The
\emph{only process} that leads to the gain in the population by
energy conservation is $\chi_1 \rightarrow \Phi\,\chi_2$ and
consequently the \emph{only} process that leads to the loss of
population with energy conservation is the inverse process of
annihilation $\Phi\,\chi_2 \rightarrow \chi_1$. The calculation of
the gain and loss terms is as follows: consider the initial Fock
state $|N_{\vk}, n^{(1)}_{\vec{p}}, n^{(2)}_{\vec{p}+\vk}\rangle $
where $N$ is the occupation of particles of $\Phi$ and $n^{(1,2)}$
that for the respective bath fields. To lowest order in the
coupling $g$ the interaction from the counterterm Lagrangian does
not contribute to the gain or loss, but only to forward scattering
since these terms are already of $\mathcal{O}(g^2)$ and the
transition probabilities will be at least of $\mathcal(g^3)$. Thus
to lowest order $\mathcal{O}(g^2)$, the gain term arises from the
following matrix element

\be \mathcal{M}_{gain}= -\frac{ig_r}{\sqrt{V}}\langle N_{\vk}+1,
n^{(1)}_{\vec{p}}-1,
n^{(2)}_{\vec{p}+\vk}+1\Big|\int_{-\infty}^{\infty}dt \,\Phi_k(t)\,
\chi_{1,\vec{p}}(t) \,\chi_{2,-\vec{p}-\vk}(t)
 \Big|N_{\vk}, n^{(1)}_{\vec{p}},
n^{(2)}_{\vec{p}+\vk}\rangle \label{matele} \ee

The limits in the time integral had been extended to $\pm \infty$
according to Fermi's Golden rule which leads to energy conservation.
The calculation of this matrix element is straightforward, taking
the absolute value squared of this matrix element, summing over
$\vec{p}$ and averaging over the occupation numbers of the particles
in the bath, which is assumed in equilibrium, one obtains the
inclusive transition probability

\be P_{gain} = \frac{2\, t}{2\Omega_k }\frac{
g^2}{32\pi^2}\Big(1+N_k\Big)\int
\frac{d^{3}\vec{p}}{\omega_{\vec{p}}^{(1)}\omega_{\vec{p}+\vec{k}}^{(2)}}\,
n(\omega_{\vec{p}}^{(1)})\,\Big[1+n(\omega_{\vec{p}+\vec{k}}^{(2)})
\Big]\,
\delta\left(\Omega_k+\omega_{\vec{p}+\vec{k}}^{(2)}-\omega_{\vec{p}}^{(1)}\right)
\ee

\noindent where $n(\omega_{\vec{p}}^{(1,2)})$ are the Bose-Einstein
distribution functions since the thermal bath is assumed to remain
in equilibrium.  To obtain the above expression we have used
$\Big|2\pi
\delta\left(\Omega_k+\omega_{\vec{p}+\vec{k}}^{(2)}-\omega_{\vec{p}}^{(1)}\right)\Big|^2=
2\pi
\delta\left(\Omega_k+\omega_{\vec{p}+\vec{k}}^{(2)}-\omega_{\vec{p}}^{(1)}\right)
t$ where $t$ is the total interaction time. A similar calculation
leads to the total transition probability for the loss process:

\be P_{loss} = \frac{2\, t}{2\Omega_k }\frac{ g^2}{32\pi^2} N_k \int
\frac{d^{3}\vec{p}}{\omega_{\vec{p}}^{(1)}\omega_{\vec{p}+\vec{k}}^{(2)}}\,
\Big[1+n(\omega_{\vec{p}}^{(1)})\Big]\,n(\omega_{\vec{p}+\vec{k}}^{(2)})
\,\delta\left(\Omega_k+\omega_{\vec{p}+\vec{k}}^{(2)}-\omega_{\vec{p}}^{(1)}\right)
\ee

The kinetic equation can now be written in the following form

\be \frac{dN_k}{dt} = (1+N_k)\,\gamma^>_k - N_k \,\gamma^<_k
\label{rateqn} \ee

The gain and loss rates  are given by

\bea \gamma^>_k & = & \frac{2 }{2\Omega_k }\frac{ g^2}{32\pi^2}\int
\frac{d^{3}\vec{p}}{\omega_{\vec{p}}^{(1)}\omega_{\vec{p}+\vec{k}}^{(2)}}\,
n(\omega_{\vec{p}}^{(1)})\,\Big[1+n(\omega_{\vec{p}+\vec{k}}^{(2)})
\Big]\,
\delta\left(\Omega_k+\omega_{\vec{p}+\vec{k}}^{(2)}-\omega_{\vec{p}}^{(1)}\right)
\label{gainrate}\\
\gamma^<_k & = & \frac{2 }{2\Omega_k }\frac{ g^2}{32\pi^2}  \int
\frac{d^{3}\vec{p}}{\omega_{\vec{p}}^{(1)}\omega_{\vec{p}+\vec{k}}^{(2)}}\,
\Big[1+n(\omega_{\vec{p}}^{(1)})\Big]\,n(\omega_{\vec{p}+\vec{k}}^{(2)})
\,\delta\left(\Omega_k+\omega_{\vec{p}+\vec{k}}^{(2)}-\omega_{\vec{p}}^{(1)}\right)\label{lossrate}
\eea

Since the bath particles are in thermal equilibrium with a
Bose-Einstein distribution function the detailed balance condition
follows, namely

\be\label{detbal}\gamma^>_k = e^{-\beta\,\Omega_k}\,\gamma^<_k \ee

The solution of the Boltzmann kinetic equation (\ref{rateqn}) is the
following

\be\label{boltz} N_k(t)=
n(\Omega_k)+[N_k(0)-n(\Omega_k)]e^{-\gamma_k\,t} \ee

\noindent where

\be\label{gama} \gamma_k= \gamma^<_k\,- \, \gamma^>_k =
\frac{2}{2\Omega_k }\, \frac{g^{2}}{32 \pi^{2}}\, \int
\frac{d^{3}\vec{p}}{\omega_{\vec{p}}^{(1)}\omega_{\vec{p}+\vec{k}}^{(2)}}\,
\Big[n(\omega_{\vec{p}+\vec{k}}^{(2)})-n(\omega_{\vec{p}}^{(1)})
\Big]
\delta\left(\Omega_k-\omega_{\vec{p}}^{(1)}+\omega_{\vec{p}+\vec{k}}^{(2)}\right)
\ee

Comparing this expression with those for the imaginary part of the
self energy given by
(\ref{imsigsplit},\ref{sig01},\ref{siga1},\ref{sigb1}) it is
straightforward to see that

\be \label{relax} \gamma_k = 2 \,\frac{\mathrm{Im}
\tilde{\Sigma}^R_r(k,\Omega_k,T)}{2\Omega_k} \ee

\noindent where

\be\label{boltgama} \mathrm{Im}\tilde{\Sigma}^R_r(k,\Omega_k, T)=
\sigma_{b,r}(k,\Omega_k,T) \ee

This expression for the relaxation rate should be compared to the
decay rate for the single \emph{quasiparticle} $\Gamma_k$ given by
eqn. (\ref{GammaT},\ref{RTTf})). Since the quasiparticle residue
in perturbation theory is $\mathcal{Z}_k(T) = 1+ \mathcal{O}(g^2)$
and the difference between the quasiparticle  frequency
$\mathcal{W}_k(T)$ and the single particle frequency $\Omega_k$ is
of $\mathcal{O}(g^2)$ to leading order in the coupling $g$, the
relaxation rate of the distribution function $\gamma_k$ and that
of the single quasiparticle $\Gamma_k$ (see eqn. (\ref{GammaT}))
is

\be \gamma_k = 2\Gamma_k +\mathcal{O}(g^4) \label{gamarate}\ee

We have provided this derivation of the usual quantum Boltzmann
equation and its solution in the case when the bath remains in
equilibrium to highlight the similarities and  differences with the
real time evolution of the distribution function given by eqn.
(\ref{finumber}):

\begin{itemize}
\item{The derivation above clearly shows that the Fock states that
enter in the matrix elements (\ref{matele}) are the asymptotic
free field Fock states associated with physical particles of mass
$m_P$. This is \emph{similar} to the definition of the
interpolating number operator (\ref{numberop}) which is based on
the free field asymptotic physical states, and includes both mass
and wave-function renormalization.  }

\item{By implementing Fermi's golden rule, namely taking the time
interval to infinity, thereby enforcing the \emph{on shell delta
function}, extracting the linear time dependence and dividing by
time to provide a \emph{local} differential equation for the time
evolution of the distribution function all memory aspects have been
neglected.  Namely implementing Fermi's golden rule results in
neglecting memory effects, which in turn results in  only
\emph{on-shell processes} contributing to the rate equation.
Contrary to this, the real time evolution of the distribution
function (\ref{finumber}) includes memory effects as is manifest in
the time integrals (\ref{funH},\ref{funF}) in (\ref{finumber}). In
turn these time integrals  keep memory of the past time evolution,
and at asymptotically long time lead to the full spectral density as
manifest in eqn. (\ref{Ninfty}), not just an on-shell delta
function. The presence of the full spectral density in the
asymptotic distribution includes  the off-shell contributions
discussed in the previous section. This discussion brings to the
fore that one of the main origins of the differences can be traced
to memory effects and the fact that the real time evolution of the
distribution function (\ref{finumber}) is \emph{non-Markovian}. The
memory of the past time evolution translates in off-shell processes
through the full spectral density. }

\item{As emphasized in   section (\ref{sec:noneLeff}) the
expression (\ref{finumber}) for the quantum kinetic  distribution
function implies a Dyson-like resummation of the perturbative
expansion and includes consistently the renormalization aspects
associated with asymptotic single particle states, namely the
correct pole mass and the wave function renormalization. The
dependence of the asymptotic distribution function on the full
spectral density is a consequence of the fluctuation-dissipation
relation.  }
\end{itemize}

\section{Conclusions and discussion}\label{sec:conc}

Motivated by a critical reassessment of the applicability of
Boltzmann kinetics in the early Universe, in this article we
studied the abundance of physical quanta of a field $\Phi$ in a
thermal plasma by introducing a quantum kinetic description based
on the non-equilibrium effective action for this field. We focused
on understanding the equilibrium abundance of \emph{particles that
are stable in the vacuum} and interact with \emph{heavier}
particles which constitute a thermal bath.

The non-equilibrium effective action is obtained by integrating
out the heavy particles to lowest order in the coupling of the
field $\Phi$ to the bath but in principle to all orders in the
coupling of the heavy fields amongst them. We show that the
non-equilibrium effective action leads to a Langevin stochastic
description with a Gaussian but colored noise and a non-Markovian
self-energy kernel. The correlation function of the noise and the
non-Markovian self-energy kernel are related by a generalized
fluctuation dissipation relation. The correlation functions are
determined by the solution of this Langevin equation which
furnishes a Dyson resummation of the perturbative expansion. We
introduced a definition of the single physical particle
distribution function in terms of a fully renormalized
interpolating Heisenberg number operator based on asymptotic
theory. The real time evolution of this single particle
distribution function is completely determined by the solution of
the Langevin equation.

We show that in a heat bath at finite temperature this number
operator becomes insensitive to the initial conditions after a time
scale $\approx 1/2\,\Gamma_k(T)$, where $\Gamma_k(T)$ is the
\emph{single quasiparticle} relaxation rate. We prove that the
asymptotic long time limit of this distribution function describes
full thermalization of the $\Phi$ particle with the thermal bath.
The equilibrium distribution function depends on  the full spectral
density and includes off-shell corrections as a result of the
non-Markovian real time evolution (with memory) and the
fluctuation-dissipation relation. Its expression is given by eqn.
(\ref{Nfinalren}). We argue that while we obtained the distribution
function in the case of a field linearly coupled to a thermal bath
of heavier particles, the final form of the distribution function at
asymptotically long time is much more generally applicable.

In order to provide a detailed assessment of novel specific features
of the distribution function in particular departure from the usual
Bose-Einstein distribution, we considered  a model in which the
thermal bath is described by two heavy bosonic fields $\chi_{1,2}$
coupled to the field $\Phi$ as $g \,\Phi\,\chi_1\,\chi_2$, with $M_1
> M_2+m_p$ and $m_p$ the pole mass of the field $\Phi$. We obtained
the real time effective action at one loop level. We find that the
in-medium processes of two body decay of the heavier particle, and
its recombination, namely $\chi_1 \leftrightarrow \chi_2 \Phi$
results in a width for the $\Phi$-particle and a broadening of its
spectral density. A detailed study of the single (physical) particle
distribution function reveals substantial corrections to the
Bose-Einstein distribution at high temperature as well as low
temperature but large momentum. At high temperature the spectral
density broadens dramatically and the off-shell contributions become
very substantial resulting in an enhancement of the abundance with
respect to the Bose-Einstein distribution. We found that at very
high temperatures, perturbation theory breaks down and a resummation
of the perturbative expansion via the renormalization group at
finite temperature may be required\cite{hotRG}. This case must be
studied further.

In the limit where the momentum of the particle is much larger than
the temperature and the masses, our analysis also reveals a
substantial departure from the Bose-Einstein distribution. In this
case the spectral density is sharply peaked near the (zero
temperature) physical pole mass, but the position of the peak moves
to higher energies. As a result, the spectral density features
off-shell contributions in a large region of frequencies
\emph{smaller} than the position of the peak. The small frequency
region is further enhanced by  temperature factors and these off
shell contributions, while exponentially small, compete with the
exponentially small  on-shell contribution which yields the
Bose-Einstein distribution. As a result the distribution function,
while strongly suppressed at high momenta  much larger than the
temperature (and mass scales), is considerably larger than the
Bose-Einstein abundance predicted by the usual Boltzmann equation.

In order to highlight the origin of the enhancement, we derived the
Boltzmann equation in \emph{renormalized perturbation theory} up to
the same order in the coupling to the bath as the non-equilibrium
effective action, which is the basis for the quantum kinetic
description. This derivation makes manifest the origin of the
discrepancy: the usual Boltzmann equation is based on Fermi's golden
rule, which requires taking a long time limit in the transition
\emph{probability}. In taking the long time limit and extracting the
asymptotic behavior of the transition probability energy
conservation is manifest as an on-shell delta function, and all
memory effects have been neglected. Furthermore in considering the
transition probability in a gain-loss balance equation, interference
phenomena have been neglected. As a result the Boltzmann equation
neglects off-shell contributions. Precisely these off-shell
contributions from the support of the spectral density away from its
peak  and near the particle mass shell, are responsible for the
departure from the Bose-Einstein result. The enhancement over the
Bose-Einstein distribution is a consequence of the  off-shell
support of the spectral density at frequencies \emph{smaller} than
the position of the peak.

Although these results were obtained within the particular specific
model studied here, the origin of the discrepancies suggests these
to be much more general. The  spectral density of  a particle that
is stable at zero temperature features an on-shell delta function
below the multiparticle thresholds. However in a medium this peak
will be broadened by different processes and the particle becomes a
quasiparticle. This unavoidable feature of an interacting particle
in a medium results in a broader spectral density with a region of
support at frequencies \emph{smaller} than the position of the peak,
which leads to a \emph{larger} contribution to the abundance as
compared to the Bose-Einstein distribution which is the ``on-shell''
result.

{\bf Cosmological consequences:}  An important feature of the
distribution function (\ref{Nfinalren}) is that it is exponentially
suppressed at low temperatures since all the intermediate states are
heavy and therefore exponentially suppressed at low temperatures.
Therefore the off-shell contributions are strongly suppressed
leading to the conclusion that the low temperature abundance is
exponentially suppressed. This is in agreement with the results of
refs.\cite{srednicki,pietroni}. Therefore we do \emph{not} expect
the low temperature enhancement of the abundance to be of any
practical relevance for cold dark matter candidates.

The consequences for the cosmic microwave background depend on the
temperature regime. For temperatures much larger than the electron
mass the photons in the plasma propagate as in-medium quasiparticles
of two species: longitudinal and transverse plasma excitations
(plasmons). The plasma frequency in the high temperature regime is
of the order $\omega_{pl} \propto \sqrt{\alpha_{em}}\,
T$\cite{kapusta,lebellac}. The corrections to the dispersion
relations (plasma frequency) arise from intermediate states of
electron-positron pairs and yield a contribution to the spectral
density with support below the light cone. These are Landau damping
processes\cite{kapusta,lebellac}  while those that yield the width
arise from Compton scattering and pair annihilation and are of
higher order. The plasmon width (up to logarithmic corrections) is
of order $\Gamma \propto \alpha^2_{em} T$. Thus the spectral
function for photons features support both above and below the light
cone, the latter is a result of Landau damping
processes\cite{kapusta,lebellac}. This latter contribution is
important because it yields support in the small frequency region
which is Bose enhanced.  Both the plasma frequency and the width
 are strong functions of temperature and we expect substantial corrections to the power
spectrum of the cosmic microwave background for $T \gg
1~\mathrm{Mev}$. However, these potential corrections are observable
only indirectly, possibly through nucleosynthesis. For temperatures
well below the electron mass the lowest order
$\mathcal{O}(\alpha_{em})$ correction to the spectral density arises
from an electron loop and an electron-positron loop (we ignore the
contribution from protons). The former gives a Landau damping cut
below the light cone
 akin to the contribution (\ref{sigb1})\cite{kapusta,lebellac} and
the latter gives a two particle cut above the pair-production
threshold. Both are off-shell contributions and yield corrections to
the spectral density which are proportional to the electron number
density ( equal to the proton number density) $n_e \sim x_e
(\Omega_b h^2_0)(1+z)^3\times 10^{-5}\mathrm{cm}^{-3}$, with $x_e$
the ionization fraction. The width of the spectral density near the
mass shell results from Compton scattering and is of order
$\alpha^2_{em}$. It is approximately given by $\Gamma \sim \sigma_T
n_e \sqrt{\frac{k_B T}{m_e}}$  and $\sigma_T$ is the Thompson
scattering cross section. During recombination the ionization
fraction diminishes precipitously within a window of redshift
$\Delta z \sim 100$ which is the width of the last scattering
surface. This rapid vanishing of the ionization fraction and
consequently of the (free) electron density entails that the
broadening of the spectrum and the spectral distortions become
vanishingly small at the end of recombination. At decoupling the
mean free path is comparable to the size of the horizon and the
spectral density for photons is basically that of free field theory.
Hence recombination erases any observable vestige of spectral
distortion through many body processes and spectral broadening, thus
there are no observable consequences of these effects in the CMB.

However we expect that our results may be potentially relevant in
the \emph{high temperature limit} for the kinetics of baryogenesis
in the Early Universe. We expect to address these issues in future
work.

 \acknowledgements  D. B.  thanks the N.S.F. for partial
support through grant PHY-0242134.

\appendix

\section{Calculation of the imaginary part of the self-energy}

The imaginary part of the self-energy is  given in the text, eqn.
(\ref{imsigrep}).

Integrating over $dp_{0}$, $dq_{0}$ and then performing the
transformation $\vec{p}\rightarrow -\vec{p}-\vec{k}$ in all the
integrals involving $n(\omega_{\vec{p}+\vec{k}}^{(2)})$, we can
write

\begin{equation}
\text{Im}\widetilde{\Sigma}^R(\omega, \vec{k})=
\sigma_{0}+\sigma_{I}+\sigma_{II}+(\sigma_{III}^{(1)}-\sigma_{III}^{(2)})
+(\sigma_{IV}^{(1)}-\sigma_{IV}^{(2)})
\end{equation}

where

\begin{eqnarray}
\sigma_{0}&=&\frac{g^{2}}{32 \pi^{2}}\, \text{sign}(\omega) \int
\frac{d^{3}\vec{p}}{\omega_{\vec{p}}^{(1)}\omega_{\vec{p}+\vec{k}}^{(2)}}\,
\delta(\,|\omega|-\omega_{\vec{p}}^{(1)}-\omega_{\vec{p}+\vec{k}}^{(2)}\,),
\\
\sigma_{I}&=&\frac{g^{2}}{32 \pi^{2}}\, \text{sign}(\omega) \int
\frac{d^{3}\vec{p}}{\omega_{\vec{p}}^{(1)}\omega_{\vec{p}+\vec{k}}^{(2)}}\,
n(\omega_{\vec{p}}^{(1)})\,
\delta(\,|\omega|-\omega_{\vec{p}}^{(1)}-\omega_{\vec{p}+\vec{k}}^{(2)}\,),
\\
\sigma_{II}&=& \frac{g^{2}}{32 \pi^{2}}\, \text{sign}(\omega) \int
\frac{d^{3}\vec{p}}{\omega_{\vec{p}}^{(2)}\omega_{\vec{p}+\vec{k}}^{(1)}}\,
n(\omega_{\vec{p}}^{(2)})\,
\delta(\,|\omega|-\omega_{\vec{p}}^{(2)}-\omega_{\vec{p}+\vec{k}}^{(1)}\,),
\\
\sigma_{III}^{(1)}&=&\frac{g^{2}}{32 \pi^{2}}\,  \int
\frac{d^{3}\vec{p}}{\omega_{\vec{p}}^{(1)}\omega_{\vec{p}+\vec{k}}^{(2)}}\,
n(\omega_{\vec{p}}^{(1)})\,
\delta(\,\omega+\omega_{\vec{p}}^{(1)}-\omega_{\vec{p}+\vec{k}}^{(2)}\,)\,
; \,\,\,\,\,
\sigma_{III}^{(2)}=\sigma_{III}^{(1)}(\omega\rightarrow -\omega),
\\
\sigma_{IV}^{(1)}&=& \frac{g^{2}}{32 \pi^{2}}\,  \int
\frac{d^{3}\vec{p}}{\omega_{\vec{p}}^{(2)}\omega_{\vec{p}+\vec{k}}^{(1)}}\,
n(\omega_{\vec{p}+\vec{k}}^{(2)})\,
\delta(\,\omega+\omega_{\vec{p}}^{(2)}-\omega_{\vec{p}+\vec{k}}^{(1)}\,)\,
; \,\,\,\,\, \sigma_{IV}^{(2)}=\sigma_{IV}^{(1)}(\omega\rightarrow
-\omega).
\end{eqnarray}

Obviously, $\sigma_{0}$ represents the zero temperature
contribution. Note that $\sigma_{II}$ and $\sigma_{IV}^{(1)}$ can be
obtained by exchanging $M_{1}$ and $M_{2}$ in $\sigma_{I}$ and
$\sigma_{III}^{(1)}$ respectively. Thus, we will only outline the
main steps in computing $\sigma_{0}$, $\sigma_{I}$ and
$\sigma_{III}^{(1)}$ in this appendix. First of all, let
$\omega_{p}=\omega_{\vec{p}}^{(1)}$ and
$z=\omega_{\vec{p}+\vec{k}}^{(2)}$. Then, we have

\begin{equation}
\sigma_{0}+\sigma_{I}=\frac{g^{2}}{16\pi k}
\text{sign}(\omega)\int_{M_{1}}^{\infty} [\,1+n(\omega_{p})\,] \,
d\omega_{p}
\int_{z^{-}}^{z^{+}}\delta(\,|\omega|-\omega_{p}-z\,)\,dz
\end{equation}

where

\begin{eqnarray}
z^{\pm}&=&\sqrt{(p\pm k)^{2}+M_{2}^{2}} \\
&=& \sqrt{\omega_{p}^{2}\pm 2k
\sqrt{\omega_{p}^{2}-M_{1}^{2}}+k^{2}-(M_{1}^{2}-M_{2}^{2})}.
\end{eqnarray}

Without loss of generality we can assume that $M_{1}>M_{2}$ for
convenience. For the integral to be non-vanishing, we require that

\begin{equation}
z^{-}<z=|\omega|-\omega_{p}<z^{+}. \label{conditionI}
\end{equation}

Squaring both sides twice properly, these two inequalities can be
reduced to $f(\omega_{p})<0$ where

\begin{equation}
f(\omega_{p})=4(|\omega|^{2}-k^{2})\omega_{p}^{2}-4|\omega|(|\omega|^{2}-a)\omega_{p}+(|\omega|^{2}-a)^{2}+4k
M_{1}^{2}
\end{equation}

and $a=k^{2}-(M_{1}^{2}-M_{2}^{2})$. Notice that the graph
$f(\omega_{p})$ against $\omega_{p}$ represents a conic with
positive y-intercept. Solving $f(\omega_{p})=0$ for $\omega_{p}$, we
obtain

\begin{equation}
\omega_{p}\equiv\omega_{p}^{\pm}=\frac{|\omega|(|\omega|^{2}-a)\pm
k\sqrt{(|\omega|^{2}-a)^{2}-4(|\omega|^{2}-k^{2})M_{1}^{2}}}{2(|\omega|^{2}-k^{2})}.
\label{omegappm}
\end{equation}

There  are two possibilities: (i) $|\omega|^{2}-k^2>0$, (ii)
$k^{2}-|\omega|^{2}>0$.  For $k^{2}-|\omega|^{2}>0$, graphs with
$f(\omega_{p})$ against $\omega_{p}$ show that condition
(\ref{conditionI}) can be satisfied only if
$\omega_{p}>\omega_{p}^{-}$ but algebraic calculation indicates that
$|\omega|-\omega_{p}^{-}<0$. Thus, condition (\ref{conditionI}) can
never be satisfied and this solution should be ignored. For
$|\omega|^{2}-k^2>0$, we have $|\omega|^{2}-a>0$.

A detailed analysis of  $f(\omega_{p})$  as well as $z^{\pm}$ and
$|\omega|-\omega_{p}$ as functions of  $\omega_{p}$ reveals that
that condition (\ref{conditionI}) can always be satisfied for
$\omega_{p}^{-}< \omega_{p}<\omega_{p}^{+}$ and
$|\omega|>\sqrt{k^{2}+M_{2}^{2}}+M_{1}$. For the discriminant in
$\omega_{p}^{\pm}$ to be positive, we require that
$|\omega|>\sqrt{k^{2}+(M_{1}+M_{2})^{2}}$ or
$|\omega|<\sqrt{k^{2}+(M_{1}-M_{2})^{2}}$. Since
$\sqrt{k^{2}+M_{2}^{2}}+M_{1}>\sqrt{k^{2}+(M_{1}-M_{2})^{2}}$, we
can only pick up $|\omega|>\sqrt{k^{2}+(M_{1}+M_{2})^{2}}$. As a
result, we conclude that

\begin{eqnarray}
\sigma_{0}&=&\frac{g^{2}}{16 \pi k}\, \text{sign}(\omega)\,
\Theta[\,|\omega|^{2}-k^{2}-(M_{1}+M_{2})^{2}\,]\,
(\omega_{p}^{+}-\omega_{p}^{-}),
\\
\sigma_{I}&=&\frac{g^{2}}{16 \pi k \beta}\, \text{sign}(\omega)
\,\Theta[\,|\omega|^{2}-k^{2}-(M_{1}+M_{2})^{2}\,]\, \ln \left(
\frac{ 1-e^{-\beta \omega_{p}^{+}} }{ 1-e^{-\beta
\omega_{p}^{-}}}\right).
\end{eqnarray}

Now, we proceed to compute $\sigma_{III}^{(1)}$:

\begin{equation}
\sigma_{III}^{(1)}=\frac{g^{2}}{16\pi k} \int_{M_{1}}^{\infty}
n(\omega_{p}) \, d\omega_{p}
\int_{z^{-}}^{z^{+}}\delta(\,\omega+\omega_{p}-z\,)\,dz.
\end{equation}

For the integral to be non-vanishing, we require that

\begin{equation}
z^{-}<z=\omega+\omega_{p}<z^{+} \label{conditionIII}
\end{equation}

which can be reduced to $g(\omega_{p})<0$ where

\begin{equation}
g(\omega_{p})=4(\omega^{2}-k^{2})\omega_{p}^{2}+4\omega(\omega^{2}-a)\omega_{p}+(\omega^{2}-a)^{2}+4k
m_{1}^{2}.
\end{equation}

Solving $g(\omega_{p})=0$ for $\omega_{p}$, we obtain

\begin{equation}
\omega_{p}\equiv\xi_{p}^{\pm}(\omega)=\frac{-\omega(\omega^{2}-a)\pm
k\sqrt{(\omega^{2}-a)^{2}-4(\omega^{2}-k^{2})M_{1}^{2}}}{2(\omega^{2}-k^{2})}
\end{equation}

First, note that $z^{\pm}\rightarrow \omega_{p}\pm k$ as $\omega_{p}
\rightarrow \infty$. Then, drawing graphs with $g(\omega_{p})$
against $\omega_{p}$ and diagrams with $z^{\pm}$ and
$\omega+\omega_{p}$ against $\omega_{p}$, we observe that condition
(\ref{conditionIII}) is always satisfied for $k^{2}-\omega^{2}>0$
with $\omega_{p}>\xi_{p}^{-}(\omega)$. For $\omega^{2}-k^{2}>0$, we
have $|\omega|^{2}-a>0$ and graphs with $g(\omega_{p})$ against
$\omega_{p}$ show that condition (\ref{conditionIII}) can be
satisfied only if $\omega<0$ and $\xi_{p}^{-}<
\omega_{p}<\xi_{p}^{+}$. Moreover, an algebraic calculation
indicates that both $\omega+\xi_{p}^{-}<0$ and
$\omega+\xi_{p}^{+}<0$ unless $\omega^{2}-k^{2}<M_{1}^{2}-M_{2}^{2}$
. Additionally, for the discriminant in $\xi_{p}^{\pm}$ to be
positive, we require that $\omega^{2}-k^{2}>(M_{1}+M_{2})^{2}$ or
$\omega^{2}-k^{2}<(M_{1}-M_{2})^{2}$. The condition
$\omega^{2}-k^{2}>(M_{1}+M_{2})^{2}$ contradicts
$\omega^{2}-k^{2}<M_{1}^{2}-M_{2}^{2}$. Hence, we must take
$\omega^{2}-k^{2}<(M_{1}-M_{2})^{2}$. Graphs of $z^{\pm}$ and
$\omega+\omega_{p}$ against $\omega_{p}$ confirm that condition
(\ref{conditionIII}) is always satisfied for
$0<\omega^{2}-k^{2}<(M_{1}-M_{2})^{2}$. As a result, we have

\begin{eqnarray}
\sigma_{III}^{(1)}-\sigma_{III}^{(2)}&=&\frac{g^{2}}{16\pi k \beta}
\,\Theta(k^{2}-\omega^{2})\ln \left( \frac{ 1-e^{-\beta
\xi_{p}^{-}(-\omega)} }{ 1-e^{-\beta
\xi_{p}^{-}(\omega)}}\right) \\
&& + \frac{g^{2}}{16\pi k \beta}\,
\text{sign}(\omega)\,\Theta(\omega^{2}-k^{2})
\,\Theta[\,k^{2}+(M_{1}-M_{2})^{2}-\omega^{2}\,] \ln \left( \frac{
1-e^{-\beta \omega_{p}^{-}} }{ 1-e^{-\beta
\omega_{p}^{+}}}\right) \\
\end{eqnarray}

where $\omega_{p}^{\pm}$ are the roots given by (\ref{omegappm}).
For $k^{2}-\omega^{2}>0$ and $\omega>0$,
$\xi_{p}^{-}(-\omega)=|\omega_{p}^{-}|$ and
$\xi_{p}^{-}(\omega)=|\omega_{p}^{+}|$. For $k^{2}-\omega^{2}>0$ and
$\omega<0$, $\xi_{p}^{-}(-\omega)=|\omega_{p}^{+}|$ and
$\xi_{p}^{-}(\omega)=|\omega_{p}^{-}|$. Therefore, we conclude that

\begin{equation}
\sigma_{III}^{(1)}-\sigma_{III}^{(2)}= \frac{g^{2}}{16\pi k
\beta}\,\text{sign}(\omega)\,\Theta[\,k^{2}+(M_{1}-M_{2})^{2}-\omega^{2}\,]
\ln \left( \frac{ 1-e^{-\beta |\omega_{p}^{-}|} }{ 1-e^{-\beta
|\omega_{p}^{+}|}}\right).
\end{equation}

Finally, to obtain $\sigma_{IV}^{(1)}-\sigma_{IV}^{(2)}$, we simply
need to exchange $M_{1}$ and $M_{2}$ in
$\sigma_{III}^{(1)}-\sigma_{III}^{(2)}$.

\end{document}